\documentclass[12pt]{paper}
\usepackage[T1]{fontenc}
\usepackage[latin9]{inputenc}
\usepackage{amsmath}
\usepackage{amsthm}
\usepackage{amssymb}
\usepackage{esint}
\usepackage{simplewick}
\usepackage{relsize}
\usepackage{alphalph}
\usepackage{enumitem}
\usepackage{chngcntr}

\usepackage[titletoc,title]{appendix}

\usepackage{tocloft}

\newcommand{\eq}[1]{\begin{align}#1\end{align}}
\newcommand{\Eq}[1]{Eq.~\eqref{#1}}

\makeatletter
\numberwithin{equation}{section}
\numberwithin{figure}{section}

\usepackage[normalem]{ulem}
\usepackage{bbm}

  \usepackage{color}
  \definecolor{darkblue}{rgb}{0.1,0.1,.7}
  \usepackage[colorlinks, linkcolor=darkblue, citecolor=darkblue, urlcolor=darkblue, linktocpage]{hyperref}

\makeatother

\begin{document}

\counterwithout{figure}{section}
\counterwithout{table}{section}

\vspace*{-2cm}

	\hfill {\tt CALT-TH-2017-041}

\vspace*{4cm}

\begin{center}
{\bf \huge TASI Lectures on \\ \bigskip Scattering Amplitudes} \\ \bigskip\bigskip\bigskip
{ \Large Clifford Cheung} \\ \bigskip\
{  \it Walter Burke Institute for Theoretical Physics \\ \smallskip California Institute of Technology, Pasadena, CA 91125} \\ \bigskip\bigskip
\end{center}

\begin{quote}

These lectures are a brief introduction to scattering amplitudes.  
We begin with a review of basic kinematical concepts like the spinor helicity formalism, followed by a tutorial on bootstrapping tree-level scattering amplitudes.  Afterwards, we discuss on-shell recursion relations and soft theorems, emphasizing their broad applicability to gravity, gauge theory, and effective field theories.   Lastly, we report on some of the new field theoretic structures which have emerged from the on-shell picture, focusing primarily on color-kinematics duality.
\end{quote}

\thispagestyle{empty}

\newpage

%\clearpage

 \setcounter{tocdepth}{2}
\tableofcontents

\thispagestyle{empty}

\newpage

\clearpage
\setcounter{page}{1}

%\address{Walter Burke Institute for Theoretical Physics \\ }

%\begin{abstract}

%These lectures are a brief introduction to scattering amplitudes.  
%We begin with a review of basic kinematical concepts like the spinor helicity formalism, followed by a tutorial on bootstrapping tree-level scattering amplitudes.  Afterwards, we discuss on-shell recursion relations and soft theorems, emphasizing their broad applicability to gravity, gauge theory, and effective field theories.   Lastly, we report on some of the new field theoretic structures which have emerged from the on-shell picture, focusing primarily on color-kinematics duality.

%\end{abstract}

\section{Introductory Remarks}

\subsection{Symmetry and Redundancy}

Quantum field theory (QFT) is a tremendously powerful tool.
It boasts applications across a range of fields, including particle physics,
string theory, cosmology, statistical mechanics, condensed matter theory, and
even economics.    In short, QFT delivers a  {\it quantitive} description of the totality of the cosmos down to the shortest distance scales probed by humankind, all while enriching you financially.

But what actually defines a QFT?  According to most textbook treatments, the answer is {\it the action}.  
 The reasons for this are practical.
First of all, the action formulation of QFT is algorithmic, automating much of the heavy lifting needed to translate high-minded theoretical structures into physical predictions.  Second, the action is borne of our classical description of the world, so it preserves---for better or worse---those natural intuitions.

Nevertheless, the luxuries of the action come at a cost, which is {\it redundancy}.    In the context of electromagnetism, this appears as the invariance of the action under gauge transformations of the photon,
\eq{
A_{\mu} & \rightarrow  A_{\mu}+\partial_{\mu}\theta.
}
The action describing interacting gravitons is similarly invariant under diffeomorphisms, which at linear order take the form,
\eq{
h_{\mu\nu} & \rightarrow  h_{\mu\nu}+\partial_{\mu}\theta_{\nu}+\partial_{\nu}\theta_{\mu}.
}
Here the transformation parameters $\theta$ and $\theta_\mu$ are {\it arbitrary} functions of spacetime with compact support.  

At face value, the appearance of an entire function's worth of freedom attests to an infinite amount of ``symmetry''.  For an apples to apples comparison, contrast this with a global Lorentz transformation, which is parameterized by six numbers for three rotations and three boosts.   This would naively suggest that gauge and diffeomorphism ``symmetries'' are infinitely more potent than their anemic global siblings.  However, as foreshadowed by the above quotation marks, this is exactly the opposite of the truth.  
Gauge and diffeomorphism invariance are actually redundancies of description created by physicists to
conveniently describe certain QFTs. 

To understand the purpose of this contrivance, recall the Wigner classification \cite{Bargmann:1948ck,Weinberg:1995mt}, which is a sort of physicist's inventory of ``what kind of stuff'' can exist in the universe even in principle.   That such a fundamental question---once thought to be squarely within the jurisdiction of philosophical navel-gazing---is answerable with absolute mathematical precision is one of the great achievements of modern physics.  In any case, the Wigner construction simply amounts to enumerating all possible irreducible unitary representations of the Poincare group.   

In four dimensions, these arguments imply that a massless particle with intrinsic spin has exactly two polarizations.  Unfortunately, our preference for manifest Lorentz invariance compels us to describe the photon and graviton with quantum fields, $A_{\mu}(x)$ and $h_{\mu\nu}(x)$, whose many indices beget four and ten degrees of freedom, respectively.  As these redundant modes are a necessary evil of manifest Lorentz covariance, we have little choice but to introduce gauge and diffeomorphism {\it redundancies} to stamp out the many degrees of freedom overzealously introduced.

The above examples suggest that this redundancy is an affliction of theories with spin.  However, it is actually {\it endemic} to the action principle.   For instance, consider a {\it scalar} field theory with the general Lagrangian\footnote{We employ mostly minus conventions where
$\eta_{\mu\nu}=\textrm{diag}(+,-,\ldots ,-)$.},
\eq{
{\cal L} & =  \frac{1}{2}K(\phi)\partial_\mu \phi  \partial^\mu \phi \label{eq:Lscalar},
}
where the field dependent kinetic term is 
\eq{
K(\phi) & = 1+ \lambda_1 \phi + \frac{1}{2!} \lambda_{2} \phi^{2}+ \frac{1}{3!} \lambda_{3} \phi^{3} + \frac{1}{4!} \lambda_{4} \phi^{4} + \ldots
}
Given the complicated and arbitrary tower of interactions defined in \Eq{eq:Lscalar}, one naturally anticipates a similarly convoluted S-matrix.  

To test this intuition, consider the simplest physical observable: the 
four-particle tree-level scattering amplitude.  A simple calculation yields
\eq{
A_4 & \propto  \sum_{i\neq j}p_{i}p_{j}\propto s+t+u=0, \label{eq:A4iszero}
}
where the usual Mandelstam invariants \cite{Mandelstam:1958xc} are  
\eq{
s &= (p_1+p_2)^2  =  (p_3+p_4)^2 \\ 
t &= (p_1+p_4)^2 = (p_2+p_3)^2 \\
u &= (p_1+p_3)^2  =(p_2+p_4)^2 .
}
Here it was absolutely crucially to impose physical external kinematics, {\it i.e.}~momenta which conserve total momentum,
\eq{
\sum_{i}p_{i} & =  0,
}
and satisfy the on-shell conditions,
\eq{
p_{i}^{2} & =  0.
}
So the four-particle amplitude actually vanishes. This effect persists---for example, the fourteen-particle amplitude also vanishes, albeit through the diabolical cancellation of upwards of five trillion Feynman diagrams\footnote{See Appendix \ref{app:count} for the counting of Feynman diagrams in a general QFT.}.

The origin of this conspiracy is that the action in \Eq{eq:Lscalar} is secretly that of a {\it free scalar}.
In particular, the S-matrix is invariant under arbitrary field redefinitions of the form \cite{Haag:1958vt,Donoghue:1992dd},
\eq{
\phi & \rightarrow f(\phi),
}
subject to the condition $f'(0)=1$ so that the weak field limit interpolates into a canonically normalized field.  For a suitably chosen function $f(\phi)$, one can then transform the free scalar action into \Eq{eq:Lscalar}.    So the S-matrix is invariant under a {\it non-symmetry} of the action!   In hindsight, this freedom of field basis should not be all that surprising.  After all, the path integral formulation of QFT treats the field as an {\it integration variable} and any integral is invariant under a proper change of variables.  

We thus arrive at the bottom line: the textbook formulation of QFT is plagued by a redundancy that undercuts any fundamental meaning to contents of the action.  By applying any old field redefinition,
we can map one action into an infinite set of {\it different}
actions describing {\it identical} physics.  

This infinity-to-one
mapping has its disadvantages.
First of all, these redundancies spawn a correspondingly redundant set of Feynman diagram representations of the same S-matrix.  For theories with gauge and diffeomorphism invariance, the equivalence of these representations is manifested by the Ward identities.  To satisfy the Ward identities the resulting Feynman diagram expressions must be immensely complicated, often exceeding the capabilities of a non-computer-assisted being even for simple physical processes.
 Second and more importantly, a poor choice of field basis may obscure or {\it altogether conceal} certain underlying structures of the theory.

\subsection{What Defines a Theory?}

So what defines a theory, if not the action?   To answer this question, consider the following Gedanken sociology experiment.  Your friend resides in another pocket of the multiverse whose entire field content is a set of massless fields which are mutually interacting. After a prolonged and ad hominem exchange about who is or is not a Boltzmann brain, they challenge you in exasperation: if you are so smart, can you compute their S-matrix
from first principles?  

For scalar fields, your first emotion might be one of relief---after all scalar Feynman diagrams are easy
to compute.  However, without any additional information, the interactions are various and sundry,
\eq{
{\cal L}_{{\rm scalar}} & =\frac{1}{2} \sum_a \partial_\mu\phi_a\partial^\mu\phi_a  + \frac{1}{3!}\sum_{abc}\omega_{abc}\phi_{a}\phi_{b}\phi_{c}+ \frac{1}{4!} \sum_{abcd}\lambda_{abcd}\phi_{a}\phi_{b}\phi_{c}\phi_d .
}
So we need to specify all of the coupling constants $\omega_{abc}$ and $\lambda_{abcd}$ to even define the action.  In other words, the theory might be simple with respect to Feynman diagrams, but it is not predictive in the slightest. 

If, on the other hand, the fields are vectors, then there is almost no freedom: the {\it only} theory permitted is Yang-Mills (YM) theory, 
\eq{
{\cal L}_{{\rm vector}} & =  -\frac{1}{4g^{2}}\sum_a F_{a\mu\nu}F_a^{\mu\nu}.
}
As we will soon see in explicit detail, this is not simply an educated guess.  One can {\it prove} that there are no interacting renormalizable theories of massless vectors aside from YM.  In this approach, the fact that the structure constants $f_{abc}$ of the gauge group are antisymmetric and satisfy the Jacobi identities is an {\it output}.

Similar logic implies that there are no self-consistent renormalizable interactions of a massless tensor. Instead, the leading allowed interactions are nonrenormalizable and coincide precisely with the Einstein-Hilbert action,
\eq{
{\cal L}_{{\rm tensor}} & =  -\frac{1}{16\pi G}\sqrt{-g}\, R .
}
So the low-energy dynamics of a massless tensor are controlled by a single parameter which is Newton's constant.

Similar statements apply to fermions.  For example, spin 1/2 fields are typically less constrained in their interactions while a spin 3/2 field can only interact consistently via supergravity \cite{McGady:2013sga}.

This little parable serves to illustrate a trivial but important point: the
simplicity of a theory hinges on how little must be specified in order to deduce its S-matrix.  Oftentimes, the mere presence of higher spin particles can be enough to fix the structure of interactions.  In some special circumstances,  this can also be true for scalars, {\it e.g.}~the pion.
As a general principle, the simplest QFTs are so constrained
that {\it were they any more constrained they would not even exist}.   

Among amplitudes practitioners, there is a growing consensus that ${\cal N}=4$ supersymmetric Yang-Mills (SYM) theory is the simplest QFT.  
To the uninitiated, this claim may appear ludicrous.  After all, ${\cal N}=4$ SYM is comprised of numerous  fields interacting through a lengthy component action.  However, this criticism is akin to bemoaning the byzantine complexity of the quantum harmonic oscillator because Hermite polynomials are hard to manipulate and there are a lot of them. Hopefully, by the end of these lectures you will understand that these naive intuitions are misguided.  A theory like ${\cal N}=4$ SYM is simple because its full perturbative S-matrix is fixed by symmetry.  The fact that the action seems complicated is merely an indication that there is a better way to compute!

The modern amplitudes program is an intellectual descendant of the fabled ``Feynman method'' of calculation, which is to {\it i}) define the problem, {\it ii}) think very hard, and {\it iii}) write down the answer while being Richard Feynman.  
Notably, mileage tends to vary in the critical last step.  A more practical approach is to construct a {\it general ansatz} for the S-matrix and sculpt out the correct answer from simple physical criteria:

\begin{itemize}[leftmargin=0.5cm,rightmargin=0.5cm]

\item \textbf{Dimensional Analysis. }

\smallskip

Scattering amplitudes should have  mass dimension consistent with the dimensionality of the coupling constants in the theory.  While this can be trivially gleaned from an action, it is still useful in practice.

\medskip

\item \textbf{Lorentz Invariance.} 

\smallskip

Scattering amplitudes should be Lorentz invariant.    For example, a four-particle amplitude of scalars is a function of $s$, $t$, $u$.  When there are particles with spin, the amplitude should also be covariant under the little group, which is a close cousin of Lorentz invariance.

\medskip

\item \textbf{Locality. }

\smallskip

Scattering amplitudes should have kinematic singularities which are consistent with factorization and unitarity. These singularities encode underlying locality of the theory.  For example, a four-particle amplitude can have poles like $1/s$ but not $1/s^2$.

\end{itemize}

\noindent  Remarkably, from these basic principles one can bootstrap a tremendous amount of physics.

\section{Crash Course in Kinematics}

Kinematics is at the heart of the modern S-matrix program.   A staple of every introductory course in QFT, this subject is typically given short shrift in the mad dash to 
dynamics.  However, as we will soon see, any sharp distinction between kinematics and dynamics is very much artificial.  Many scattering amplitudes are actually {\it uniquely} fixed by kinematic constraints like Lorentz invariance and momentum conservation.

In the textbook formulation of QFT, kinematic data is characterized more or less exclusively by momentum vectors,
\eq{
p_{\mu} & =  (p_{0},\vec{p}),
}
where throughout these lectures we take the convention that all particles are {\it incoming}. 
Over the years, the study of scattering amplitudes program has revealed a
 menagerie of alternative variables: spinor helicity \cite{DeCausmaecker:1981jtq,Kleiss:1985yh,Xu:1986xb,Gunion:1985vca} 
 , twistors \cite{Penrose:1967wn}, and momentum
twistors \cite{Hodges:2009hk}, just to name a few.    Of course, these all describe the same underlying physics.   However, by recasting expressions into the appropriate variables, one can achieve massive simplifications which reveal otherwise invisible structures.
In short, never bring a knife to a gun fight
because they have different symmetries.

\subsection{Spinor Helicity}

Spinor helicity variables are a tremendously powerful tool for representing on-shell kinematics.   While our discussion here will center exclusively on spinor helicity formalism in four dimensions, generalizations exist for lower \cite{Agarwal:2008pu} and higher \cite{Cheung:2009dc,CaronHuot:2010rj} dimensions.

In a nutshell, the spinor helicity formalism maps the components of a four-vector into those of a two-by-two matrix via
\eq{
p_{\alpha\dot{\alpha}} & =  p_{\mu}\sigma^{\mu}_{\alpha\dot{\alpha}}=\left(\begin{array}{cc}
p_{0}+p_{3} & p_{1}-ip_{2}\\
p_{1}+ip_{2} & p_{0}-p_{3}
\end{array}\right),
}
where $\sigma^{\mu}  =  (1,\vec \sigma)$ is a four-vector of Pauli matrices and
the undotted and dotted indices transform under the usual spinor representations of the Lorentz group.    
The only Lorentz invariant quantity which can be constructed from $p_{\alpha \dot\alpha}$ is its determinant,
\eq{
\textrm{det}(p)=p^{\mu}p_{\mu}=0, \label{eq:onshell}
}
which vanishes for a massless on-shell particle.

The spinor helicity formalism has been generalized to include massive particles \cite{Dittmaier:1998nn,Schwinn:2006ca,Boels:2009bv}.   However, the honest truth is that theories with massless particles are unambiguously simpler than their massive counterparts.  This fact transcends the context of scattering amplitudes.   Even in the action formulation, particles like gluons, gravitons, chiral fermions, or Nambu-Goldstone bosons (NGBs) are all massless as a consequence of symmetries which also stringently constrain their permitted interactions.  This is the reason why we so often define massive particles as a {\it controlled perturbation} away from the massless limit, {\it e.g.}~in the case of soft breaking of supersymmetry or chiral symmetry.

Returning to \Eq{eq:onshell}, we see that $p_{\alpha\dot{\alpha}}$ has vanishing determinant.  Since $p_{\alpha\dot{\alpha}}$ is  nonzero, it is a two-by two matrix of at most rank one, which without loss of generality can be written as the outer product of two two-component
objects which are called spinors,
\eq{
p_{\alpha\dot{\alpha}}=\lambda_{\alpha}\tilde{\lambda}_{\dot{\alpha}}. 
}
Note that $\lambda_{\alpha}$
and $\tilde{\lambda}_{\dot{\alpha}}$ are sometimes called ``holomorphic'' and ``anti-holomorphic'' spinors because of their transformation properties under the Lorentz group.  Bear in mind that there is no connection here to any underlying
fermionic states---these spinors are {\it not} anti-commuting
Grassmann numbers, but rather plain old complex numbers.  

For real momenta, $p_{\alpha \dot\alpha} $ is Hermitian, implying the reality condition,
\eq{
\tilde{\lambda}_{\dot\alpha}= \pm (\lambda^{*})_{\dot{\alpha}},
}
for positive and negative energy states, respectively.
When the momentum is complex,
there is no relation between $\lambda_{\alpha}$ and $\tilde{\lambda}_{\dot{\alpha}}$ and they are independent.  However, in this case $p_{\alpha \dot\alpha}$ still satisfies \Eq{eq:onshell} so these spinors still describe a massless particle.
Strictly speaking, since the external momenta in physical processes are real, we should maintain this reality condition throughout our calculations.  However, as we will soon see, much of our power will flow from
analytic continuation to complex momenta while treating the physical
S-matrix as a restriction of this more general object.  

We are now ready to introduce the Lorentz invariant building blocks of the spinor helicity formalism.  Given two particles $i$ and $j$, we define
\eq{
\langle ij\rangle & =  \lambda_{i\alpha}\lambda_{j\beta}\epsilon^{\alpha\beta}\\
{}[ij] & =  \tilde{\lambda}_{i\dot{\alpha}}\tilde{\lambda}_{j\dot{\beta}}\epsilon^{\dot{\alpha}\dot{\beta}} ,
}
which are commonly referred to as ``angle'' and ``square'' brackets.  Any function of four-dimensional kinematic data can be written exclusively in terms of these objects.    For example, angle and square brackets come together to form the familiar
Mandelstam invariants, 
\eq{
s_{ij}&=(p_{i}+p_{j})^{2}  =  2p_{i}p_{j}=\langle ij\rangle[ij],
}
and likewise for higher-point kinematic invariants.

Because spinors are simple objects, they are subject to relatively few algebraic manipulations.   One important property is antisymmetry, $\langle ij\rangle = -\langle ji\rangle$ and $[ij] = -[ji]$, which in turn implies that $\langle ii\rangle=[ii]=0$.  Moreover, since each spinor is two-dimensional, one can always write a spinor as a linear combination of two linearly independent spinors, 
\eq{
\langle ij\rangle  \lambda_k+   \langle ki\rangle \lambda_j+  \langle jk\rangle \lambda_i& =  0,
}
which is known as the Schouten identity.

In point of fact, spinor helicity variables are nothing more than an algebraic reshuffling of the external kinematic data.   Such a manipulation would not be particularly advantageous were it not for the fact that scattering amplitudes enjoy an  {\it immense} reduction in complexity when translated into these variables.  The avatar of this simplification is the S-matrix for gluon scattering.    Depending on the level of sadism in your introductory QFT course, it is quite likely that you have computed the four-gluon amplitude and possibly even the five-gluon amplitude but not higher.    The amplitude for six-particle scattering was completed in a brute-tour-de-force Feynman diagram calculation of Parke and Taylor \cite{Parke:1985ax}.  The latter involves two hundred and twenty diagrams, spelled out in several pages of nested variables and tables of numerical coefficients.  Shortly after, Parke and Taylor brilliantly realized that for a particular ``maximally helicity violating'' (MHV) helicity configuration, {\it i.e.} two negative helicity gluons with the rest positive, this mountain of algebra telescopes into an expression built from simple color structures multiplying {\it monomial} expressions of the form
\eq{
A( \cdots i^- \cdots j^- \cdots) = \frac{\langle ij\rangle^4}{\langle 12\rangle \langle 23\rangle \cdots \langle n1\rangle}, \label{eq:PT}
}
which is the celebrated Parke-Taylor formula \cite{Parke:1986gb}.

So what is the moral of this particular story?  First, we see with brutal clarity that almost all of the terms in the Feynman diagram expansion are nothing more than a {\it complicated rewriting of zero}.  Second, new features of the amplitude become manifest: in particular, the Parke-Taylor formula is purely a function angle brackets, revealing an underlying holomorphic structure of MHV amplitudes.   As observed long ago by Nair \cite{Nair:1988bq}, this property led in part to the twistor-string description of gluon scattering \cite{Witten:2003nn} as well as more recent developments connecting soft gluon radiation to an underlying conformal field theory \cite{He:2015zea,Cheung:2016iub}.

\subsection{Little Group}

Spinor helicity variables have a certain aesthetic elegance.  But why are they practically useful? As is so often the case, their utility flows from the fact that they linearly realize the symmetries of the system.  These symmetries are Lorentz invariance and the so-called little group, which we now discuss. 

The little group is comprised of the subset of Lorentz transformations which leave the momentum $p_{\mu}$ of a particle invariant.  According to the Wigner classification \cite{Bargmann:1948ck,Weinberg:1995mt}, a particle of momentum $p_\mu$ is then described by an irreducible representation of the little group.
 For example, consider the momentum vector of a massless particle in four dimensions,
\eq{
p_{\mu} & =  (p_0,0,0,p_0),
}
shown here in a certain frame.  The transformations which leave $p_\mu$ invariant form the $ISO(2)$ group of translations and rotations
in the plane transverse to the trajectory of the particle. The finite dimensional representations of
this group are eigenvectors of the rotation subgroup.  The corresponding eigenvalues label the helicities of physical states.

Under the little group, the spinor helicity variables $\lambda_{i}$ and $\tilde{\lambda}_{i}$ transform so as to leave $p_{i}$ is unchanged, so
\eq{
\lambda_{i}  \rightarrow  t_{i}\lambda_{i}\qquad \textrm{and} \qquad
\tilde{\lambda}_{i} & \rightarrow  t_{i}^{-1}\tilde{\lambda}_{i}.
}
Note that for
real momenta, the spinors satisfy a reality condition so for physical kinematics $t_{i}$ is restricted to be a pure phase.

Of what use is the little group if it leaves all momenta
$p_{i}$ invariant?   For interacting scalars, the answer is {\it of no use at all}, since these particles are singlets of the little group and the amplitude is solely a function of $p_i$.   On the other hand, for particles with spin it is necessary to introduce nontrivial representations of the little group which carry helicity quantum numbers.  In terms of conventional Feynman
diagrams, this little group covariance enters through an additional set
of kinematic objects we have thus far ignored: the polarizations. 

In terms of spinor helicity variables, polarization vectors take the form
\eq{
e_{\alpha\dot{\alpha}}^{+}  =  \frac{\eta_{\alpha}\tilde{\lambda}_{\dot{\alpha}}}{\langle\eta\lambda\rangle}\qquad \textrm{and}\qquad
e_{\alpha\dot{\alpha}}^{-}  =  \frac{\lambda_{\alpha}\tilde{\eta}_{\dot{\alpha}}}{[\tilde \lambda \tilde \eta]}, \label{eq:polarization_def}
}
where the $\pm$ superscripts label helicity.  By construction, these polarizations are transverse
to the momenta, so
\eq{
p^{\alpha\dot{\alpha}}e_{\alpha\dot{\alpha}}^{+}  \propto [\tilde{\lambda}\tilde{\lambda}]=0 \qquad \textrm{and} \qquad p^{\alpha\dot{\alpha}} e_{\alpha\dot{\alpha}}^{-}& \propto  \langle {\lambda}{\lambda}\rangle =0.
}
Note the appearance of ``reference spinors'', $\eta$ and $\tilde{\eta}$, which are 
 linearly independent
of $\lambda$ and $\tilde{\lambda}$ but otherwise arbitrary.  In fact, one can even assign {\it different} 
reference spinors for each external particle.   How is this possible?  If we change $\eta$
and $\tilde{\eta}$, should not the answer change?  To see why this yields no contradiction, consider substituting the reference spinor $\eta$ with a different but still arbitrary reference spinor $\eta'$,
\eq{
\eta_{\alpha} & \rightarrow  \eta_{\alpha}'=a\eta_{\alpha}+b\lambda_{\alpha},
}
where we have expanded $\eta'$ as a linear
combination of the original reference spinor $\eta$ and the momentum
spinor $\lambda$.  
The corresponding polarization vector is similarly shifted according to
\eq{
e_{\alpha\dot{\alpha}}^{+} & \rightarrow  e_{\alpha\dot{\alpha}}^{+\prime}= %\frac{(a\eta+b\lambda)_{\alpha}\tilde{\lambda}_{\dot{\alpha}}}{(a\langle\eta|+b\langle\lambda|)|\lambda\rangle}=
e_{\alpha\dot{\alpha}}^{+}+\left(\frac{b}{a\langle\eta\lambda\rangle}\right)p_{\alpha\dot{\alpha}},
}
which is literally a gauge transformation.  
So the polarizations {\it do} change depending on the choice of reference but in a way that leaves the amplitude invariant by the Ward identity.   

Under the little group, the polarization vectors transform as
\eq{
e_{\alpha\dot{\alpha}}^{+}  \rightarrow t^{-2}e_{\alpha\dot{\alpha}}^{+} \qquad \textrm{and} \qquad
e_{\alpha\dot{\alpha}}^{-}  \rightarrow t^{2}e_{\alpha\dot{\alpha}}^{-},
}
so the little group covariance of each polarization encodes its associated helicity.  A scattering amplitude with multiple particles will then be multilinear in the corresponding polarizations, so
\eq{
A(1^{h_1} \cdots n^{h_n})& =  e_{\mu_{1}}^{h_{1}}\ldots e_{\mu_{n}}^{h_{n}}A^{\mu_{1}\cdots\mu_{n}}, \label{eq:eA}
}
where $h_{i}=\pm1 $ labels the helicity of each leg.   Hence, the scattering amplitude is little group covariant with weight
\eq{
A(1^{h_1} \cdots n^{h_n}) & \rightarrow  \prod_{i}t_{i}^{-2h_{i}}A(1^{h_1} \cdots n^{h_n}). \label{eq:little_group_covariance}
}
At first glance it may seem peculiar that the scattering amplitude is covariant rather than invariant under the little group.  After all, all physical observables are invariant.  Nevertheless, there is no contradiction because on real kinematics  the little group parameters $t_{i}$ reduce to pure phases which trivially cancel in the matrix element squared. 

The tensorial object on the right-hand side of \Eq{eq:eA}, $A^{\mu_{1}\cdots\mu_{n}}$, is the travesty computed from Feynman diagrams in the standard approach.  Meanwhile, the object on left-hand side, $A(1^{h_1} \cdots n^{h_n})$, labels ``helicity amplitudes,'' which are a set of distinct functions corresponding to each helicity configuration.   These objects are {\it true} scattering amplitudes in the sense that they are fully gauge invariant functions like the Parke-Taylor formula in \Eq{eq:PT}.
  The aim of the amplitudes program is to sidestep $A^{\mu_{1}\cdots\mu_{n}}$ altogether by bootstrapping $A(1^{h_1} \cdots n^{h_n})$ directly from the physical principles outlined previously.

Reference spinors can be tremendously useful for explicit calculations.  As a simple example, consider the amplitude for all plus helicity gluons, $A(1^+ 2^+ \cdots n^+)$.  First, note that every term in the Feynman diagram expansion appears with at least one factor of $e_i^+ e_j^+$.  This is because the amplitude has mass dimension $4-n$ and at most $n-3$ propagators.  Thus, the numerator of every  Feynman diagram has $n-2$ powers of momenta or fewer, implying that at least two polarization vectors are contracted via $e_i^+ e_j^+$.  However, by choosing the {\it same} reference spinor, $\eta$, for all polarizations, we set $e_i^+ e_j^+ =0$, establishing that $A(1^+ 2^+ \cdots n^+)=0$.  A similar argument applies to the all but one plus helicity gluon amplitude, $A(1^- 2^+ \cdots n^+)$.  Again, every term enters with at least one factor of $e_i^+ e_j^+$, so choosing $\eta = \lambda_1$ for every reference sets $e_i^+ e_j^+ = e_1^- e_i^+ =0$ so $A(1^- 2^+ \cdots n^+)=0$ as well.  Without even doing a calculation, we have proven that the leading nontrivial tree-level gluon scattering amplitudes is MHV.

\section{Bootstrapping Amplitudes}

The goal of this section is to systematically enumerate all possible Lorentz invariant interactions among massless particles in four dimensions.  In accordance with our stated philosophy these results will be derived without reference to an action.    This concept of an ``amplitudes bootstrap'' closely follows the seminal work of Benincasa and Cachazo \cite{Benincasa:2007xk}, though similar treatments can be found in the existing review literature \cite{Elvang:2015rqa,Zee:2003mt}.

\subsection{Three-Particle Amplitudes}

The leading nontrivial contribution to the S-matrix is the three-particle amplitude.  According to momentum conservation, 
\eq{
p_{1}+p_{2}+p_{3}=0  \qquad\Rightarrow\qquad  \begin{array}{c}
(p_{1}+p_{2})^{2}=\langle12\rangle[12]=p_{3}^{2}=0\\
(p_{2}+p_{3})^{2}=\langle23\rangle[23]=p_{1}^{2}=0\\
(p_{3}+p_{1})^{2}=\langle31\rangle[31]=p_{2}^{2}=0
\end{array}
}
From the top line, we deduce that if $\langle12\rangle\neq0$, then $[12]=0$.  Furthermore, $\langle12\rangle[23]=-\langle 11 \rangle[13] -\langle 13 \rangle[33] =0$, so $[23]=0$. Repeating this procedure cyclically, we find that $[12]=[23]=[31]=0$.  If, on the other hand, we assume $[12]\neq 0$, then $\langle 12\rangle = \langle 23\rangle = \langle 31\rangle=0$.
In summary, the three-particle amplitude only has support on two possible kinematic configurations: holomorphic, corresponding to all vanishing square brackets,
\eq{
  [12]=[23]=[31]=0  \qquad \Rightarrow\qquad\tilde{\lambda}_{1}\propto\tilde{\lambda}_{2}\propto\tilde{\lambda}_{3},% \quad(\textrm{holomorphic}),
}
and anti-holomorphic, corresponding to all vanishing angle brackets,
\eq{
  \langle12\rangle=\langle23\rangle=\langle31\rangle=0  \qquad \Rightarrow\qquad\lambda_{1}\propto\lambda_{2}\propto\lambda_{3} . %\quad(\textrm{anti-holomorphic}).
}
Both kinematic configurations imply that $p_1 p_2 = p_2 p_3 = p_3 p_1=0$ and require complex momenta so that $\lambda_i$ and $\tilde \lambda_i$ are independent variables.

Without loss of generality, the three-particle amplitude in the holomorphic kinematic configuration takes the form
\eq{
A(1^{h_{1}}2^{h_{2}}3^{h_{3}})=\langle12\rangle^{n_{3}}\langle23\rangle^{n_{1}}\langle31\rangle^{n_{2}}. \label{eq:A3ansatz}
}
Imposing the criterion of little group covariance in \Eq{eq:little_group_covariance}, we solve for the exponents  $n_{i}$ in terms of the helicities $h_i$ of the external particles,
\eq{
\begin{array}{c}
-2h_{1}=n_{2}+n_{3}\\
-2h_{2}=n_{3}+n_{1}\\
-2h_{3}=n_{1}+n_{2}
\end{array}\qquad\Rightarrow\qquad\begin{array}{c}
n_{1}=h_{1}-h_{2}-h_{3}\\
n_{2}=h_{2}-h_{3}-h_{1}\\
n_{3}=h_{3}-h_{1}-h_{2}
\end{array}
}
Note that the exponents are integers because all helicities $h_i$ are integers or half-integers and fermions always appear in pairs.

The mass
dimension of the three-particle amplitude cannot be negative, as this would require nonlocality due to inverse powers of derivatives in the three-particle vertex.  Hence, the assumption of locality implies that the three-particle amplitude has nonnegative mass dimension, so 
\eq{
[A(1^{h_{1}}2^{h_{2}}3^{h_{3}})]=n_{1}+n_{2}+n_{3}=-(h_{1}+h_{2}+h_{3})=-h\geq0 , \label{eq:sumh}
}
where $[\ldots]$ denotes the mass dimension accumulated by powers of momenta, ignoring the intrinsic dimensionality of coupling constants.  From \Eq{eq:sumh} we see that holomorphic kinematics applies for $h\leq0$ while anti-holomorphic
kinematics applies for $h\geq0$.  
In summary, we have derived a general formula for the three-particle amplitude of massless particles in four dimensions,
\eq{
A(1^{h_{1}}2^{h_{2}}3^{h_{3}})  = 
\left \{
\begin{array}{ll}
 \langle12\rangle^{h_{3}-h_{1}-h_{2}}\langle23\rangle^{h_{1}-h_{2}-h_{3}}\langle31\rangle^{h_{2}-h_{3}-h_{1}}  ,\quad & h\leq 0 \\
 {[}12{]}^{ h_{1}+h_{2}-h_{3}}   {[}23{]}^{h_{2}+h_{3}-h_{1}} {[}31{]}^{h_{3}+h_{1}-h_{2}}   ,\quad & h\geq 0  
 \end{array}
 \right.
%A_{3}^{{\rm MHV}}(1^{h_{1}}2^{h_{2}}3^{h_{3}}) & = & \langle12\rangle^{(h_{3}-h_{1}-h_{2})}\langle23\rangle^{(h_{1}-h_{2}-h_{3})}\langle31\rangle^{(h_{2}-h_{3}-h_{1})}\qquad\qquad\,(h\leq0)\\
%A_{3}^{{\rm \overline{{\rm MHV}}}}(1^{h_{1}}2^{h_{2}}3^{h_{3}}) & = & [12]^{-(h_{3}-h_{1}-h_{2})}[23]^{-(h_{1}-h_{2}-h_{3})}[31]^{-(h_{2}-h_{3}-h_{1})}\qquad(h\geq0)
\label{eq:A3}
}
which like the tuxedo t-shirt is both practical and elegant.

Since this formula was derived purely from symmetry and dimensional analysis, it holds {\it nonperturbatively}.   For example, loops which would naively generate logarithms or more complicated functional objects like $\log(p_{1}p_{2})$ are all singular or vanishing on three-particle kinematics.

In the subsequent sections, we will study \Eq{eq:A3} for the familiar cases of scalars, vectors, and tensors.

\subsubsection{Scalars}

For identical scalars, all helicities $h_{i}=0$ are vanishing and \Eq{eq:A3} implies that the three-particle amplitude is a constant,
\eq{
A(123) & =  \omega,
}
corresponding to a single scalar self-interacting through a cubic potential term.   A similar result applies to multiple scalars, 
\eq{
A(1_{a} 2_{b} 3_{c} ) & =  \omega_{a b c} ,\label{eq:A3scalar}
}
where the subscripts run over flavors.  Since the external states are bosons, the coupling constant $\omega_{abc}$ is symmetric in its indices.

We have argued that \Eq{eq:A3scalar} is the only possible on-shell three-particle amplitude of massless scalars.  But what about derivatively coupled scalars, \emph{e.g.}~interacting through $\phi(\partial\phi)^{2}$?  In fact, these amplitudes vanish on-shell because they can only depend on products of momenta and these necessarily vanish: $p_{1}^{2}=p_{2}^{2}=p_{3}^{2}=p_{1}p_{2}=p_{2}p_{3}=p_{3}p_{1}=0$.   At the level of the action, this triviality can be made manifest by applying a field redefinition that simply eliminates these cubic derivative interactions altogether.

\subsubsection{Vectors}

For the case of identical vectors the helicities are $h_i = \pm 1$.  This implies that the exponents $n_i$ in \Eq{eq:A3ansatz} are odd integers.   Consequently, the three-particle amplitude is {\it odd} under the exchange of any two external states---even though they are bosons.  Self-consistency then demands that the three-particle amplitude must identically {\it vanish}.  This conclusion is in perfect agreement with the textbooks: the three-particle amplitude of photons is indeed zero by charge conjugation symmetry.

While this argument forbids cubic self-interactions of a single vector, it still allows for interactions among multiple species---also known as colors---of vectors.  Introducing a color index to each vector, we obtain 
\eq{
A(1_{a}^{-}2_{b}^{-}3_{c}^{+})=f_{abc}\frac{\langle12\rangle^{3}}{\langle13\rangle\langle32\rangle} \qquad \textrm{and}\qquad   A(1_{a}^{+}2_{b}^{+}3_{c}^{-})=f_{abc}\frac{[12]^{3}}{[13][32]}. \label{eq:AYM}
}
Here $f_{abc}$ has to be fully antisymmetric in its indices so that the amplitude is {\it even} under exchange of bosons.  
These expressions exactly agree with the more complicated expressions obtained from Feynman diagrams,
\eq{
A(1_a 2_b 3_c ) =
%e_1^{\mu_1} e_2^{\mu_2} e_3^{\mu_3}A_{\mu_1 \mu_2 \mu_3} =
f_{abc} (e_1 e_2 )(p_1 e_3 - p_2 e_3) + \textrm{cyclic perm},
}
after inserting the expressions for the polarization vectors in  \Eq{eq:polarization_def}.

For the all minus and all plus helicity combinations, \Eq{eq:A3} reduces to
\eq{
A(1_{a}^{-}2_{b}^{-}3_{c}^{-})=f_{abc}\langle12\rangle\langle23\rangle\langle31\rangle, & \qquad  A(1_{a}^{+}2_{b}^{+}3_{c}^{+})=f_{abc}[12][23][31], \label{eq:Fcubed_amp}
}
where $f_{abc}$ is again antisymmetric so the amplitude is even under exchange of bosons. 
By counting powers of momentum, it is clear that \Eq{eq:Fcubed_amp} originates from a three-derivative interaction among gluons.   This arises from a {\it nonrenormalizable} higher dimension operator, $f_{abc} F_{\mu}^{a\nu}F_{\nu}^{b\rho}F_{\rho}^{c\mu}$, which is generated from loops of heavy colored particles.

\subsubsection{Tensors}

For the case of identical tensors the helicities are $h_i = \pm 2$.  From \Eq{eq:A3ansatz}, we see that the exponents $n_i$ are even, so the amplitude is invariant under the exchange of bosons.  The resulting three-particle amplitude is 
\eq{
A(1^{--}2^{--}3^{++})=\frac{\langle12\rangle^{6}}{\langle13\rangle^{2}\langle32\rangle^{2}}, & \qquad & A(1^{++}2^{++}3^{--})=\frac{[12]^{6}}{[13]^{2}[32]^{2}}, \label{eq:A3grav}
}
corresponding to the scattering of three gravitons.  As before, we can compare this expression to the equivalent Feynman diagram expression.   Even with the aid of a simplified action for graviton perturbations \cite{Cheung:2016say}, we obtain
\eq{
A(123)=
-\frac{1}{2}\left(e_1 e_2\right){}^2
\left(p_1 e_3\right)
   \left(p_2 e_3\right)+ \left(e_1 e_2\right)\left(e_2
   e_3\right)
   \left(p_1 e_3\right) \left(p_2
   e_1\right)  + \textrm{perm},
   }
%\eq{
%A(123)=&
%-\left(e_1\cdot e_2\right){}^2
%\left(p_1\cdot e_3\right)
%   \left(p_2\cdot e_3\right)+\left(e_2\cdot
%   e_3\right) \left(e_1\cdot e_2\right)
%   \left(p_1\cdot e_3\right) \left(p_2\cdot
%   e_1\right)
 %  \nonumber\\
%   &+\left(e_1\cdot e_3\right)
%   \left(e_1\cdot e_2\right) \left(p_1\cdot
%   e_2\right) \left(p_2\cdot
%   e_3\right)+\left(e_2\cdot e_3\right)
%   \left(e_1\cdot e_2\right) \left(p_2\cdot
%   e_3\right) \left(p_3\cdot
%   e_1\right)
%   \nonumber\\
%   &+\left(e_1\cdot e_3\right)
%   \left(e_1\cdot e_2\right) \left(p_1\cdot
%   e_3\right) \left(p_3\cdot
%   e_2\right)+\left(e_1\cdot e_3\right)
%   \left(e_2\cdot e_3\right) \left(p_1\cdot
%   e_2\right) \left(p_3\cdot
%   e_1\right)
%   \nonumber\\
%   &-\left(e_2\cdot e_3\right){}^2
%   \left(p_2\cdot e_1\right) \left(p_3\cdot
%   e_1\right) -\left(e_1\cdot e_3\right){}^2
%   \left(p_1\cdot e_2\right) \left(p_3\cdot
%   e_2\right)
%   \nonumber\\
%   &+\left(e_1\cdot e_3\right)
%   \left(e_2\cdot e_3\right) \left(p_2\cdot
%   e_1\right) \left(p_3\cdot e_2\right),
%   }
which is significantly more complicated then \Eq{eq:A3grav}. Note that here and for the remainder of these lectures, all tensor polarizations will be written as products of vector polarizations.

Last but not least, consider the amplitude for tensors scattering in the all minus and all plus configurations,
\eq{
A(1^{--}2^{--}3^{--})=\langle12\rangle^2 \langle23\rangle^2 \langle31\rangle^2 , & \qquad  A(1^{++}2^{++}3^{++})=[12]^2[23]^2[31]^2.
}
Again counting powers of momentum, we see that these amplitudes are generated by a six-derivative interaction among gravitons induced by the curvature-cubed operator, $R_{\mu\nu}^{\;\; \;\;\rho\sigma}R_{\rho\sigma}^{\;\; \;\;\alpha\beta}R_{\alpha\beta}^{\;\; \;\;\mu \nu}$.   

It is peculiar that we have jumped so quickly to {\it cubic} curvature invariants.  After all, what happened to the {\it quadratic} curvature invariants, $R^2$, $R_{\mu\nu}R^{\mu\nu}$, and $R_{\mu\nu\rho\sigma}R^{\mu\nu\rho \sigma}$?  By explicit calculation, one can verify that these operators produce demonstrably nontrivial Feynman vertices.  Have we somehow missed the four-derivative contributions to the amplitude?

On the contrary, the above calculations are perfectly sound.  Moreover, they actually foreshadow one of the great surprises in the history of quantum gravity.  Here we refer to the pioneering work of 't Hooft and Veltman \cite{tHooft:1974toh}, who  first calculated  the one-loop divergences of four-dimensional pure gravity.  By naive power counting, one-loop graviton diagrams should produce divergent contributions corresponding to local curvature-squared counterterms like $R^2$, $R_{\mu\nu}R^{\mu\nu}$ and $R_{\mu\nu\rho\sigma}R^{\mu\nu\rho \sigma}$.  
There is, however, a huge caveat.  None of these operators actually contribute to on-shell scattering amplitudes in four dimensions!  You can verify this explicitly---the off-shell Feynman diagrams are complicated and non-trivial but their effects evaporate on-shell.  

To understand this at the level of the action, rewrite $R_{\mu\nu\rho\sigma}R^{\mu\nu\rho \sigma}$ in terms of the Gauss-Bonnet combination, $
R^2 - 4 R_{\mu\nu} R^{\mu\nu} + R_{\mu\nu\rho\sigma}R^{\mu\nu\rho\sigma}$.  In four dimensions this is a total derivative which decouples from perturbative scattering.  Exploiting the freedom of field redefinitions, then plug the leading order equations of motion---that is, Einstein's field equations---into the next-to-leading order corrections.  This sets $R=R_{\mu\nu}=0$ into the curvature-squared operators, eliminating them altogether.  

The upshot here is twofold:  in four dimensions, curvature-squared operators do not contribute to the tree-level graviton S-matrix, and pure gravity is {\it finite} at one-loop.  This second fact briefly prompted optimism that pure gravity might be finite at higher orders---a flighting dream soon crushed by a two loop divergence \cite{Goroff:1985th,Goroff:1985sz,vandeVen:1991gw,Bern:2015xsa} corresponding to the $R_{\mu\nu}^{\;\; \;\;\rho\sigma}R_{\rho\sigma}^{\;\; \;\;\alpha\beta}R_{\alpha\beta}^{\;\; \;\;\mu \nu}$ counterterm.  

Nowadays, the divergence structure of supergravity theories remains an active field of study.   Explicit calculations have exposed unexpected cancellations not anticipated by counterterm arguments \cite{Bern:2012cd,Bern:2012gh,Bossard:2012xs, Bossard:2013rza,Bern:2013qca,Bern:2014sna}, suggesting the  enticing possibility that ${\cal N}=8$ supergravity might actually be ultraviolet finite to {\it all orders}.  These developments call to mind the timeworn adage, ``all that is not forbidden is compulsory.'' Though true at a buffet, such wisdom must be applied with care to ${\cal N}=8$  supergravity.

\subsection{Four-Particle Amplitudes}

That the three-particle amplitude is wholly fixed by symmetry is perhaps unsurprising.  After all, these objects are generated by pure contact vertices and do not suffer from the complications of exchanged particles.   For the same reason, extending our analysis to four-particle amplitudes will require a new ingredient: locality.  

Locality is encoded in the singularity structure of an amplitude.  In particular, consider when the external kinematics are carefully tuned so that an intermediate particle is on-shell.  The tree-level amplitude exhibits a $1/p^2$ singularity signalling the propagation of the intermediate state over a macroscopic physical distance in spacetime.   In this factorization limit the amplitude decomposes into a product of subamplitudes corresponding to two distinct processes occurring at disparate points.  

This factorization property places stringent conditions on the structure of the amplitude.  For example, in the four-particle amplitude, simple poles like $1/s$ can arise, but not poles like $1/s^2$.   For the tree-level four-particle amplitude, this implies that
\eq{
\underset{s\rightarrow0}{\textrm{lim}}\,s\,A_{4} & =  A_{3}\,A_{3},
}
which is a highly nontrivial relationship linking higher-point and lower-point
kinematics.  As a trivial corollary, the mass dimensions of the four-particle and three-particle amplitudes are related by
\eq{
 & [A_{4}]=2[A_{3}]-2 \label{eq:massdim4},
}
where as before, our definition of mass dimension tabulates contributions from kinematic factors but not couplings.

In the remainder of this section, we derive the four-particle tree amplitudes for massless particles purely from factorization and our results for three-particle amplitudes.  

\subsubsection{Scalars}

As we saw earlier, the three-particle amplitude has positive mass dimension,
$[A_{3}]\geq0$, since the associated vertex should be local.  By dimensional analysis, \Eq{eq:massdim4} then implies that
\eq{
[A_{3}]\geq0 & \qquad\Rightarrow\qquad  [A_{4}]\geq-2, \label{eq:A4scalar_bound}
}
so the mass dimension of the four-particle amplitude of scalars is bounded from below.  Furthermore, the tree amplitude should be a permutation invariant function of $s$, $t$, and $u$ with only simple poles.  Enumerating all possible such functions, we obtain
\eq{
\phi^{3}\quad  \textrm{:} & \quad  A_{4}=s^{-1} +t^{-1} +u^{-1}\\
\phi^{4}\quad \textrm{:} & \quad  A_{4}=1\\
(\partial\phi)^{2}\phi^{2}\quad \textrm{:} & \quad  A_{4}=s+t+u=0\\
(\partial\phi)^{4}\quad \textrm{:} & \quad  A_{4}=s^{2}+t^{2}+u^{2}\\
(\partial\partial\phi)^{2}(\partial\phi)^{2}\quad \textrm{:} & \quad  A_{4}=s^{3}+t^{3}+u^{3},
}
and so on.
Note that the $\phi^{3}$ amplitude has singularities in every factorization
channel.  In contrast, all the derivatively coupled scalar amplitudes are regular, consistent with the fact that their associated on-shell three-particle amplitudes are zero according to our previous arguments.  

\subsubsection{Vectors}

Next, consider the four-particle tree amplitude of massless vectors.  From our earlier analysis we saw that $[A_{3}]=1$, so \Eq{eq:massdim4} implies that
\eq{
[A_{3}]=1 & \qquad\Rightarrow\qquad  [A_{4}]=2[A_{3}]-2=0,
}
and the amplitude is dimensionless.  

By incorporating the constraints of little group covariance
and dimensional analysis, it is possible to build a restrictive ansatz for the four-particle amplitude.   To do so, we exploit that amplitude transforms as
\eq{
A(1_{a}^{-}2_{b}^{-}3_{c}^{+}4_{d}^{+}) & \rightarrow  t_{1}^{2}t_{2}^{2}t_{3}^{-2}t_{4}^{-2}A(1_{a}^{-}2_{b}^{-}3_{c}^{+}4_{d}^{+}),
}
under the little group.  Factoring out the full little group weight of the amplitude, we consider the general ansatz,
\eq{
A(1_{a}^{-}2_{b}^{-}3_{c}^{+}4_{d}^{+}) & =  \langle12\rangle^{2}[34]^{2}F(s,t,u),
}
where we have defined
\eq{
F(s,t,u) & =  \frac{c_{st}}{st}+\frac{c_{tu}}{tu}+\frac{c_{us}}{us},
}
which is the most general little group invariant function of proper mass dimension and with only simple poles.  By dimensional analysis, the coefficients
$c_{st},c_{tu},c_{us}$ are dimensionless constants.

Demanding factorization on the $s$-channel singularity, we find that
\eq{
&\underset{s\rightarrow0}{\textrm{lim}} \,s\,A(1_{a}^{-}2_{b}^{-}3_{c}^{+}4_{d}^{+})  =  \langle12\rangle^{2}[34]^{2}\frac{1}{t}\left(c_{st}-c_{us}\right) \nonumber \\
 & =  \sum_{h=\pm}\sum_{e}A(1_{a}^{-}2_{b}^{-}P_{e}^{h})A(3_{c}^{+}4_{d}^{+}P_{e}^{-h})=\sum_{e}A(1_{a}^{-}2_{b}^{-}P_{e}^{+})A(3_{c}^{+}4_{d}^{+}P_{e}^{-})\nonumber \\
 & =  \sum_{e}f_{abe}f_{cde}\frac{\langle12\rangle^{3}}{\langle1P\rangle\langle P2\rangle}\frac{[34]^{3}}{[3P][P4]} =  \sum_{e}f_{abe}f_{cde}\langle12\rangle^{2}[34]^{2}\frac{1}{t} ,\label{eq:A4vector_fact}
}
where we have defined $P = -(p_1+ p_2)  = (p_3+p_4)$ and set $t=-u$ on the factorization channel.
In the last line we have used that $\langle1P\rangle[P4]=-\langle12\rangle[24]$
and $\langle P2\rangle[3P]=\langle42\rangle [34]$. 

Comparing the first and last lines of \Eq{eq:A4vector_fact}, we obtain an equation for $c_{st}-c_{us}$ in terms of the structure constants.
Repeating this exercise for the $t$- and $u$-channels
we obtain a triplet of equations,
\eq{
c_{st}-c_{us} & =  \sum_{e}f_{abe}f_{cde}\\
c_{tu}-c_{st} & =  \sum_{e}f_{bce}f_{ade}\\
c_{us}-c_{tu} & =  \sum_{e}f_{cae}f_{bde},
}
whose sum is an equation with a very old name,
\eq{
\sum_{e}f_{abe}f_{cde}+ f_{bce}f_{ade}+f_{cae}f_{bde} & =  0,
}
{\it i.e.}~the Jacobi identity.  In the textbook approach to YM theory, the Jacobi identity is a byproduct of an underlying Lie algebra---nothing more than an algebraic identity, ${\rm tr}([T_{a},T_{b}][T_{c},T_{d}])+{\rm tr}([T_{b},T_{c}][T_{a},T_{d}])+{\rm tr}([T_{c},T_{a}][T_{b},T_{d}])=0$.  Here the Jacobi identity is a {\it consistency condition} arising from factorization of the four-particle amplitude! 

\subsubsection{Tensors}

Last but not least is the four-particle tree amplitude of massless tensors.  From our earlier analysis we learned that the mass
dimension of the three-particle graviton amplitude is $[A_{3}]=2$, so \Eq{eq:massdim4} implies that
\eq{
[A_{3}]=2 & \qquad\Rightarrow\qquad  [A_{4}]=2[A_{3}]-2=2.
}
Like before, we construct a general ansatz for the four-particle amplitude
consistent with little group covariance and dimensional analysis,
\eq{
A(1^{--}2^{--}3^{++}4^{++}) & =  \langle12\rangle^{4}[34]^{4}F(s,t,u),
}
where $F(s,t,u)$ is the unique little group and permutation invariant object of the correct mass dimension with
only simple poles,
\eq{
F(s,t,u) & \propto  \frac{1}{stu}.
}
The above formulas agree with the known expression for the four-particle graviton amplitude.  Note that we have arrived at this result  without even imposing factorization.

\subsection{Diverse Dimensions}

Thus far we have restricted to four dimensions in order to reap the benefits of the spinor helicity formalism.  As one might expect, however, an analogous version of the bootstrap can be implemented in $d$ dimensions.   Starting from an ansatz constructed from $d$-dimensional kinematic invariants, one imposes consistency conditions like gauge invariance and factorization in order to excise  the physical scattering amplitude.  

Admittedly, this approach is an unabashed retreat from our stated principles.  After all, we had vowed earlier to banish the notion of gauge invariance from all future discussions.  Nevertheless, the following exercise gives a flavor of the mechanics of the amplitudes bootstrap in diverse dimensions.

To begin, consider the three-particle amplitude.  The complete basis of independent Mandelstam invariants for three particles is\footnote{See Appendix \ref{app:mandelstam} for a derivation of this basis for any number of particles.}
\eq{
%I^{(3)} &=
\left\{ 
p_1 e_2, p_1 e_3, p_2 e_1,
e_1 e_2, e_1 e_3, e_2e_3 \right\}.
}
For the case of gluons, the amplitude is linear in $e_1$, $e_2$, and $e_3$, so the most general ansatz with the power counting of YM theory is
\eq{
A(1_a 2_b 3_c ) =c_1 (e_1 e_3) (p_1 e_2) + c_2 (e_1 e_2) (p_1 e_3) + c_3 (e_2 e_3)(p_2 e_1) .
}
By applying the Ward identity to each external leg, one can verify that the only combination of terms that satisfies the Ward identities is proportional to $(c_1,c_2,c_3) \propto (-1,1,1)$,
which agrees with the expression from Feynman diagrams as well as the result in \Eq{eq:AYM}.

For gravitons, the amplitude is instead quadratic in $e_1$, $e_2$, and $e_3$, so the most general ansatz with the power counting of gravity is
\eq{
A(123) = &
    \phantom{{}+{}} d_1 \left(e_1 e_3\right){}^2
   \left(p_1 e_2\right){}^2
   +d_2
   \left(e_1 e_2\right)
   \left(e_1 e_3\right)
   \left(p_1 e_3\right)
   \left(p_1 e_2\right) \nonumber \\
   &+d_3
   \left(e_1 e_2\right){}^2
   \left(p_1 e_3\right){}^2
   +d_4
   \left(e_1 e_3\right)
   \left(e_2 e_3\right)
   \left(p_2 e_1\right)
   \left(p_1 e_2\right)
    \nonumber \\
   &+d_5
   \left(e_1 e_2\right)
   \left(e_2 e_3\right)
   \left(p_1 e_3\right)
   \left(p_2 e_1\right)
   +d_6
   \left(e_2 e_3\right){}^2
   \left(p_2 e_1\right){}^2.
}
Similarly, one can check that the only diffeomorphism invariant combination is proportional to $(d_1,d_2,d_3,d_4,d_5,d_6) \propto (1, -2, 1, -2, 2, 1)$, which again matches Feynman diagrams and our expression derived in \Eq{eq:A3grav}.

It is straightforward to extend these arguments to higher point by including factorization as an additional input.  We will not go through this exercise here.  However, it has been shown \cite{Arkani-Hamed:2016rak,Rodina:2016mbk,Rodina:2016jyz} that an even milder set assumptions is actually sufficient to derive the S-matrix.  In fact, one can generate the full tree-level S-matrices of YM and gravity using only gauge invariance and certain weak assumptions about the singularity structure.

Note that the graviton three-particle amplitude is {\it literally} the square of the gluon three-particle amplitude in any dimension.  At face value this is absolutely miraculous, though anticipated by the four-dimensional expressions derived in \Eq{eq:AYM} and \Eq{eq:A3grav}.  Nevertheless, given our bootstrapping approach it should be obvious that this relationship was nearly {\it inevitable}.   In particular, since the three-particle amplitude of gluons is a gauge invariant function, so too is its square.  This squared object matches the power counting of gravity, so it is automatically a viable candidate function for the three-particle amplitude of gravitons.  
Hence, the only real surprise is that there are no other gauge invariant objects {\it beyond this} and the graviton amplitude is unique.

\section{Recursion Relations}

The previous sections describe a first principles derivation of all possible three- and four-particle scattering amplitudes.  While these methods straightforwardly generalize to higher-point, there is thankfully a less artisanal approach.  In particular, on-shell recursion {\it automates} this factorization procedure by rewriting on-shell amplitudes in terms of lower-point on-shell amplitudes.   When the entire tree-level S-matrix is defined recursively from a finite number of seed amplitudes, the corresponding theory is said to be {\it on-shell constructible}.    Pioneered in the seminal work of Britto, Cachazo, Feng, and Witten \cite{Britto:2004ap,Britto:2005fq}, on-shell recursion has since blossomed into a veritable cottage industry.

At first glance, on-shell recursion relations share a superficial similarity with Feynman diagrams.  After all, every Feynman diagram is computed as a sum over products of subdiagrams.  In fact, Berends and Giele long ago derived compact {\it off-shell} recursion relations \cite{Berends:1987me} which efficiently repackage the Feynman diagrammatic expansion.   Despite this, on-shell recursion offers the important conceptual benefit of a purely {\it on-shell} definition of the S-matrix.  In this picture, the off-shell ``bulk'' spacetime associated with the action formalism is demonstrably unnecessary.

In what follows, we derive on-shell recursion relations for scattering amplitudes in a broad range of QFTs.  Our discussion will be light on explicit examples, focusing more on the broad implications for QFT in general.  For the avid reader seeking the gory details, we suggest the excellent treatments of on-shell recursion in the review literature \cite{Elvang:2015rqa,Zee:2003mt}.

\subsection{Momentum Shifts}

On-shell recursion is a systematic procedure for relating an amplitude to its values at singular kinematics.  In order to probe these kinematic configurations we define a momentum shift, which is a one-parameter deformation of the
external momenta engineered to sample various kinematic limits.  Here we will consider shifts of the form
\eq{
p_{i} & \rightarrow p_i(z) =  p_{i}+zq_{i} \label{eq:momentum_shift},
}
where $z$ is a complex number parameterizing deviations from the original kinematic configuration.  Note that some of the $q_i$ may be zero if there are unshifted external legs.  The shifted momenta in turn induce a deformation of the amplitude,
\eq{
A & \rightarrow  A(z).
}
 It is crucial that the momentum shift preserve on-shell kinematics for {\it all}  $z$, so the deformed amplitude $A(z)$ is still perfectly physical despite its gruesome moniker.  Hence, the momentum shifts are restricted by total momentum conservation,
\eq{
\sum_{i}p_{i}(z)  =  0 \qquad \Rightarrow \qquad
\sum_{i}q_{i}  =  0, \label{eq:shift_conditions_mom}
}
together with the on-shell conditions,
\eq{
p_{i}(z)^2  =0 \qquad \Rightarrow \qquad q_{i}^{2}=q_{i}p_{i}  =  0.\label{eq:shift_conditions_onshell}
}
\Eq{eq:shift_conditions_mom} and \Eq{eq:shift_conditions_onshell} should be interpreted as conditions on $q_i$.  These equations are generally easy to satisfy.

\subsection{Factorization from Analyticity}

At last we are ready to derive a general formula for on-shell recursion.  To begin, we exploit the trivial fact that the original amplitude, $A(0)$, is obtained from the residue of $A(z)$ at $z=0$,
\eq{
A(0) & =  \frac{1}{2\pi i}\underset{z=0}{\oint}dz\;\frac{A(z)}{z} 
= -\sum_{I}\underset{z=z_{I}}{\textrm{Res}}\left[\frac{A(z)}{z}\right]+B_{\infty} \label{eq:preRR},
}
where we have used Cauchy's theorem to express the residue at $z=0$ in terms of the residues at all other poles, $z_I$, plus a boundary contribution, $B_\infty$.  Crucially, this manipulation is only permitted if $A(z)$ is a {\it rational} function of the kinematics, {\it i.e.}~the amplitude is tree-level.  Also, note that
$B_{\infty}=0$ whenever $A(z)$ vanishes at large $z$.   

Each pole in $z_I$ labels a kinematic singularity of the amplitude.  On physical grounds, this {\it necessarily} coincides with a physical factorization channel.    To see this explicitly, consider a subset $I$ of external particle carrying total momentum,
\eq{
P_{I}=\sum_{i\in I}p_{i}  \qquad\Rightarrow\qquad  P_{I}(z)=\sum_{i\in I}p_{i}(z), 
}
after the momentum shift.  Here the shifted total momentum is
\eq{
P_I(z) = P_I + z Q_I,
}
where $Q_I = \sum_{i\in I}q_{i}$.  For general kinematics,  $P_{I}^{2}\neq0$ and so  $P_{I}$ is off-shell. However, the deformed momentum $P_I(z)$ {\it is} on-shell when
\eq{
0=P_{I}^{2}(z_{I}) & =  P_{I}^{2}+2z_{I}Q_I P_{I}\qquad \Rightarrow \qquad z_{I}=-\frac{P_{I}^{2}}{2Q_I P_{I}}.
}
For simplicity we have assumed that the $q_i q_j=0$ so that $Q_I^2=0$ for any subset of legs $I$. Strictly speaking this is not mandatory but the calculation simplifies substantially when $P_I^2(z)$ is a linear function.  For the more general case of $Q_I^2 \neq0$,  $P_I^2(z)$ is quadratic in $z$ and contributions from this factorization channel will be split between two singular points corresponding to the roots of this polynomial \cite{Cheung:2015cba}.

At the point $z_I$, the amplitude should factorize into a product of lower-point amplitudes,
\eq{
\underset{z\rightarrow z_{I}}{\textrm{lim}}P_{I}^{2}(z)A(z) & =  A_{L}(z_{I})A_{R}(z_{I}),
}
which implies that the associated residue is
\eq{
\underset{z=z_{I}}{\textrm{Res}}\left[\frac{A(z)}{z}\right] & =  \frac{1}{z_{I}}\left(\frac{dP_I^{2}(z_{I})}{dz_{I}}\right)^{-1}\times\underset{z\rightarrow z_{I}}{\textrm{lim}}\left[P_I^{2}(z)A(z)\right]\\
% & = & \left(-\frac{2qP_{I}}{P_{I}^{2}}\right)\left(\frac{1}{2qP_{I}}\right)\times A_{L}(z_{I})A_{2}(z_{I})\\
 & =  -\frac{1}{P_{I}^{2}}\times A_{L}(z_{I})A_{R}(z_{I}).
}
Plugging back into \Eq{eq:preRR}, we obtain our final expression
\eq{
A(z=0) & =  \sum_{I}\frac{1}{P_{I}^{2}}A_{L}(z_{I})A_{R}(z_{I})+B_{\infty}, \label{eq:RR}
}
which is a general formula for on-shell recursion.  Here $I$ runs over all singularities $z_I$ in the amplitude.  

While each $z_I$ pole corresponds to a factorization channel, the converse may not be true.  In fact, the only factorization channels which enter into \Eq{eq:RR} are those that have shifted momenta on {\it both} sides of the channel.  If instead all the shifted momenta were on {\it one} side of the channel, then $P_I(z) = P_I$ and there would be no value of $z$ for which $P_I^2(z)=0$ and thus no associated singularity.   The fact that \Eq{eq:RR} may only contain a subset of factorization channels should not trouble you in the least---irrespective of the choice of momentum shift, Cauchy's theorem guarantees that the final result will be the correct  amplitude.  On the contrary, by shifting as few legs as possible we obtain a recursion relation involving fewer factorization channels, thus resulting in a more compact expression for the amplitude.

\subsection{Boundary Terms}

The recursion relation in \Eq{eq:RR} recasts {\it any} tree-level on-shell amplitude in terms of lower-point on-shell amplitudes and a boundary term, $B_\infty$, whose value hinges crucially on the form of the momentum shift.
Much like human beings, momentum shifts are organized into two distinct
and easily discernible types: good ($B_{\infty}=0$) and bad ($B_{\infty}\neq0$).  For a good momentum shift, the corresponding amplitude is constructible purely from lower-point on-shell data.  Furthermore, we say that a theory is on-shell constructible if there exists a momentum shift for which $B_{\infty}=0$ for every tree-level scattering amplitude\footnote{In some circumstance one can relate $B_\infty$ to other physical data \cite{Feng:2009ei}, though we will not consider this possibility here.}.  As we will soon see, this is criterion is satisfied in an enormous class of theories.

There exist several methods for evaluating the boundary term.  The first \cite{Britto:2005fq} is to simply use a known representation of the amplitude, {\it e.g.}~the MHV expansion \cite{Cachazo:2004kj}.
A second approach involves a return to the action.  Here the key insight is to realize that the large $z$ limit coincides with a particular high energy limit where the shifted legs behave like hard particles scattering in a background of soft quanta \cite{ArkaniHamed:2008yf,Cheung:2008dn}.  To calculate $B_\infty$ one then computes the amplitude for hard particle scattering in the background field method.  A third way---more in line with the underlying philosophy of the amplitudes program---is to derive $B_\infty$ purely from general principles like dimensional analysis, Lorentz invariance, and locality \cite{Cohen:2010mi}.

Before continuing, let us comment briefly on {\it why} it is that certain amplitudes are not on-shell constructible.  The classic example of this is the four-particle amplitude of $\phi^4$ theory, $A(1234) = \lambda$.  Because this amplitude is momentum independent it will not vanish at large $z$ for {\it any} momentum shift.  Naively, this sounds like a spectacular failure: on-shell methods cannot even build the simplest imaginable amplitude!  

However, such a criticism is tremendously unfair because it holds on-shell recursion relations to a standard not held even by Feynman diagrams.  In particular, $A(1234)$ arises from a Feynman vertex which is simply {\it assumed} in the action formalism.  That is, the $\phi^4$ vertex is primordial and unrelated by symmetry to lower order operators.  Hence, $A(1234)$ is not related to the three-particle amplitude by factorization and should be included as a {\it seed} for the on-shell recursion.  On-shell recursion can then be used to construct all amplitudes with   greater than four particles in $\phi^4$ theory.   The general lesson here is that interaction vertices which are unrelated by symmetry to lower-order processes are not on-shell constructible and this should be completely unsurprising.

\subsection{BCFW Recursion}

The first incarnation of on-shell recursion, pioneered by BCFW \cite{Britto:2004ap,Britto:2005fq}, remains to this day unparalleled in its simplicity.    BCFW defined a shift of two external legs given by \Eq{eq:momentum_shift} where
\eq{
q_1=-q_2  =  \lambda_{1}\tilde{\lambda}_{2} \label{eq:BCFW_shift}.
}
In terms of spinor helicity variables, this shift is elegantly written as
\eq{
\tilde{\lambda}_{1} & \rightarrow  \tilde{\lambda}_{1}+z\tilde{\lambda}_{2}\\
\lambda_{2}& \rightarrow \lambda_{2}-z\lambda_{1} \label{eq:BCFW_shift},
}
which automatically preserves \Eq{eq:shift_conditions_mom} and \Eq{eq:shift_conditions_onshell}.
Since the momentum shift only involves particles 1 and 2, the sum over $I$ in \Eq{eq:RR} only involves factorization channels in which particles 1 and 2 appear on opposite sides of the channel.

Remarkably, the boundary term $B_\infty$ vanishes both in gauge theory and gravity as long as the momentum shift is applied to external legs with the appropriate configuration of helicities \cite{Britto:2005fq,ArkaniHamed:2008yf}.   Presented here without proof, the boundary term is
\eq{
A(1^{-}2^{-}\cdots),A(1^{-}2^{+}\cdots), A(1^{+}2^{+}\cdots)  &\quad\Rightarrow\quad  B_{\infty}=0\\
A(1^{+}2^{-}\cdots)  &\quad\Rightarrow\quad  B_{\infty}\neq0, \label{eq:B_YM}
}
depending on the helicity configuration of particles 1 and 2.  At first glance it may seem peculiar that $B_{\infty}$ is not invariant under the exchange of plus and minus helicities.  However, this is purely an artifact of  \Eq{eq:BCFW_shift}, which picks out a chirality between particles 1 and 2.

Fortunately, in the case of YM theory it is always possible to choose a set of adjacent legs whose helicities match one of the good momentum shifts in \Eq{eq:B_YM}.  This is true because every nontrivial tree-level amplitude in YM theory has at least one plus leg and one minus leg.  We thus conclude that YM theory is on-shell constructible via BCFW recursion.

A corollary of this result is that the tree-level S-matrix of YM theory can be built purely from the three-particle amplitudes we derived in \Eq{eq:AYM}.  This should come as a shock---after all, the YM quartic term is critically necessary in order to maintain  gauge invariance of the action.  However, on-shell recursion plainly demonstrates that this term carries {\it zero} physical content beyond the cubic interaction vertex.

 Remarkably, the BCFW recursion relations are also applicable to graviton scattering amplitudes \cite{Benincasa:2007qj}.  In this case the boundary term is 
\eq{
A(1^{--}2^{--}\cdots),A(1^{--}2^{++}\cdots),A(1^{++}2^{++}\cdots)  &\quad\Rightarrow\quad  B_{\infty}=0\\
A(1^{++}2^{--}\cdots)  &\quad\Rightarrow\quad  B_{\infty}\neq0.
}
The fact that graviton amplitudes vanish at large $z$ for any momentum shift is actually quite remarkable given the infamously poor ultraviolet behavior of gravity.  Despite these intuitions, gravity is closely analogous to gauge theory and similarly on-shell constructible. 

Just as in YM theory, the tree-level graviton S-matrix can be reduced via on-shell recursion to the primordial three-particle amplitudes we derived earlier in \Eq{eq:A3grav}.  Here we learn that the {\it infinite tower} of higher order graviton vertices literally serve no purpose but to manifest diffeomorphism invariance at the level of the action.

\subsection{All-Line Recursion}

We could have foreseen that YM theory and gravity are on-shell constructible.  After all, both theories have highly constrained interactions controlled by a single coupling constant.   What {\it is} surprising is that on-shell constructibility is a property of {\it any renormalizable QFT in four dimensions}.  Conveniently, this tremendously large class of theories also happens to have real world relevance.

It is possible to derive on-shell recursion relations for a generic QFT using an all-line holomorphic momentum shift \cite{Elvang:2008vz,Cohen:2010mi},
\eq{
q_i =  c_i \eta \tilde \lambda_i, \label{eq:all_line_q}
}
where $\eta$ is an arbitrary reference spinor and $c_i$ are constants with each external leg.  This is a generalization of the three-line momentum shift devised by Risager \cite{Risager:2005vk} to prove the MHV vertex expansion \cite{Cachazo:2004kj}.  
Like the reference spinors used to define external polarization vectors, $\eta$ will appear in intermediate steps in the recursion but will cancel in any physical answers. By writing \Eq{eq:all_line_q} as a shift of the holomorphic spinors,
\eq{
\lambda_i \rightarrow \lambda_i + z c_i \eta, \label{eq:all_line_lambda}
}
we see that \Eq{eq:shift_conditions_onshell} is automatically satisfied while \Eq{eq:shift_conditions_mom} implies 
\eq{
\sum_i^n c_i \tilde \lambda_i =0. \label{eq:all_line_cond}
}
Meanwhile, an anti-holomorphic momentum shift is defined in the obvious way by swapping the holomorphic and anti-holomorphic spinors in \Eq{eq:all_line_q}, \Eq{eq:all_line_lambda}, and \Eq{eq:all_line_cond}.

With nothing more than little group covariance and dimensional analysis \cite{Cohen:2010mi}, one can show that the large $z$ behavior of an all-line shift is
\eq{
\lim_{z\rightarrow \infty} A_n(z)  = z^m,
}
where the exponent of the fall off is
\eq{
m &= \left\{
\begin{array}{ll}
 \frac{1}{2}(4-n-[g]+h), &\quad \textrm{holomorphic shift}\\
 \frac{1}{2}(4-n-[g]-h), &\quad \textrm{anti-holomorphic shift}
\end{array}
\right. 
}
Here $h$ is the total helicity of the external particles and $[g]$ denotes the mass dimension of all the coupling constants in the amplitude. For the case of a renormalizable theory, $[g]\geq 0$, so
\eq{
m &\leq  \left\{
\begin{array}{ll}
 \frac{1}{2}(4-n+h), &\quad \textrm{holomorphic shift}\\
 \frac{1}{2}(4-n-h), &\quad \textrm{anti-holomorphic shift}
\end{array}
\right.
}
For $n>4$, the boundary term in \Eq{eq:RR} vanishes for holomorphic shifts when $h\leq0$ and for anti-holomorphic shifts when $h\geq 0$.  Since this accounts for every possibility, all amplitudes with $n>4$ particles are on-shell constructible in a renormalizable QFT.  Amplitudes with $n\leq 4$ are the seed amplitudes for this recursion.

In practice, all-line recursion is substantially more complicated to implement than BCFW.  Since all the momenta are shifted, the corresponding recursion relation will involve contributions from {\it every} factorization channel of the amplitude.   Furthermore, the resulting expressions carry spurious dependence on the reference spinor $\eta$, thus obscuring the underlying Lorentz invariance of the final answer. 

On the other hand, all-line recursion offers the conceptual victory of elevating on-shell methods to the same footing as the tree-level action for all renormalizable QFTs.   Admittedly, these results are confined to  four dimensions but recall that so too are many of your close friends and family.  That is to say, these methods provide a purely on-shell definition of our actual laws of physics, {\it i.e.}~the standard model plus gravity.

\section{Infrared Structure}

The previous sections detail how physical properties like dimensional analysis, Lorentz invariance, and locality dictate the tree-level S-matrices of an enormous class of theories.  Clearly, our methods have the most horsepower in theories either involving renormalizable interactions or particles with spin.  Notably missing from this list are {\it nonrenormalizable} theories of scalars, including numerous effective field theories (EFTs) which are critically important for low-energy phenomena.  The archetypal example of an EFT is the nonlinear sigma model (NLSM) \cite{GellMann:1960np,Coleman:1969sm,Callan:1969sn}, which famously describes the universal dynamics of spontaneous symmetry breaking.  This theory plays an important role in the strong interactions \cite{Weinberg:1968de} and electroweak physics \cite{Appelquist:1980vg,Longhitano:1980iz}.  

Unfortunately, there are significant obstacles to extending on-shell methods to scalar EFTs.  First of all, scalars are singlets of the little group, so the associated considerations are simply not useful for constraining the amplitude. Moreover, the inherent nonrenormalizability of EFTs exacerbates the issue of boundary terms in on-shell recursion.  Of course, this second concern is not completely warranted since the same is true of gravity, where  we have seen that poor high energy behavior is unimportant.

A greater impediment is that scalar EFTs simply require {\it more} physical input than just dimensional analysis, Lorentz invariance, and locality.  In the textbook formulation of EFTs, the additional ingredient is nonlinearly realized symmetry, {\it e.g.}~chiral symmetry in the NLSM.  Unlike gauge and diffeomorphism invariance, this is {\it genuine} global symmetry with physical consequences for the S-matrix.  

Unfortunately, the concepts of nonlinearly realized symmetry, coset spaces, and related paraphernalia are inextricably tied to the notion of an action.  Unmoored from these cherished principles, how can we possible construct the S-matrix of an EFT?  As we will see, the path forward is to introduce additional physical information which is admissible in the language of on-shell scattering amplitudes: the soft limit.

The soft limit of an amplitude is defined by sending the momentum
of an external leg to zero.  To characterize the behavior in the soft limit, we define a soft degree $\sigma$ \cite{Cheung:2014dqa,Cheung:2016drk},
\eq{
\underset{p\rightarrow0}{\textrm{lim}} \, A(p) & \propto  p^{\sigma},
}
which is integer-valued for any tree-level amplitude.  When the right-hand side of this equation involves a structure which is universal, {\it e.g.}~zero or a known factor times a lower-point amplitude, it is said that the amplitude satisfies a soft theorem.  

There is a long prehistory of deriving soft theorems from an action using chiral symmetry or gauge and diffeomorphism invariance.  From this perspective the soft theorems are a {\it byproduct} of the action.  In the subsequent sections, we implement an amusing inversion of this logic: by {\it assuming} the soft theorems we can derive the S-matrix of the EFT and hence all of its properties.  This is a notable reversal of the usual reductionist approach to high energy physics.   Eschewing a top-down perspective  grounded in symmetry and unification in the ultraviolet, we pursue a bottom-up definition of these theories gleaned from the deep infrared.

\subsection{Weinberg Soft Theorems}

The prototypical soft theorem was discovered in the seminal work of Weinberg \cite{Weinberg:1964ew,Weinberg:1965nx} on gauge theory and gravity.  As we have shown, since the these theories are on-shell constructible by dimensional analysis, Lorentz invariance, and factorization, these soft theorems cannot encode any additional information beyond this.  Nevertheless, the case of gauge theory and gravity will serve as an instructive warmup.

Gauge theory and gravity amplitudes are actually {\it singular} in the soft limit because soft particles can interact with hard particles through a primordial cubic vertex. Since the soft particle imparts negligible momentum, the hard particle propagator adjacent to the interaction is nearly on-shell, yielding a singularity.  Hence, the leading contribution to the soft limit of gauge theory and gravity enters at $\sigma = -1$.    

First, let us consider first the case of quantum electrodynamics (QED).  As proven by Weinberg \cite{Weinberg:1965nx}, the soft limit of a photon of momentum $p_\mu$ and polarization $e_\mu$ is
\eq{
\lim_{p\rightarrow 0} \, A_{n+1} = e_\mu S^\mu A_n,
}
where $S^\mu$ is a universal soft function that is {\it independent} of the hard process.  On general grounds we know that every term in $S^\mu$ should have a propagator pole together with one power of momentum from the cubic gauge interaction vertex.  Indeed, the Weinberg soft factor is
\eq{
S^\mu &=  \sum_{i}^n q_i \frac{ p_i^\mu}{p_i p},
}
where $q_i$ is the coupling constant associated with the cubic vertex attaching to each hard particle.  By the Ward identity, the soft theorem should be invariant under $e_{\mu} \rightarrow e_{\mu} + \alpha p_\mu $, so 
\eq{
\left( \sum_{i}^n q_i\right) A_n =0.
}
Here $q_i$ should be interpreted as the electric charge of each hard particle, whose sum is conserved by the soft theorem\cite{Weinberg:1964ew}. 

An analogous story holds for gravitons, 
\eq{
\lim_{p\rightarrow 0} \, A_{n+1} =  e_{\mu\nu}  S^{\mu\nu} A_n.
}
In this case, every term in $S^{\mu\nu}$ has a propagator pole together with two powers of momenta from the cubic vertex with the graviton, yielding
\eq{
S^{\mu\nu} = \sum_{i}^n  \kappa_i\frac{ p^\mu_i p^\nu_i}{ p_i p} .
}
The soft theorem should be invariant under a diffeomorphism transformation, $e_{\mu\nu} \rightarrow e_{\mu\nu} + \alpha_\mu p_\nu + \alpha_\nu p_\mu $, so we obtain
\eq{
\left( \sum_{i}^n \kappa_i p_i^\mu\right) A_n =0.
}
We are already familiar with an equation of this form---total momentum conservation implies that $\sum_{i}^n p_i^\mu =0$.  Any additional constraint on the external momenta would then require that the {\it hard} particles in the amplitude conform to some special kinematic configuration.  As this is plainly impossible, we deduce that 
\eq{
\kappa_i = \kappa,
}
so the coupling strength of gravity is {\it universal} \cite{Weinberg:1964ew}.   Thus, we have arrived at the celebrated equivalence principle of Einstein, derived here without the ethical pitfalls of hurling unsuspecting test subjects into the dead of space in an elevator.

As one might anticipate, there exist soft theorems for gauge theory and gravity beyond leading order in the soft momentum.  In particular, terms which are finite in the soft limit also exhibit various universal structures, {\it e.g.}~for the subleading soft photon theorem, \cite{Low:1958sn,Burnett:1967km} and the subleading \cite{Gross:1968in,Jackiw:1968zza,White:2011yy,Cachazo:2014fwa}
and subsubleading soft graviton theorems \cite{Cachazo:2014fwa}.

\subsection{Adler Zero and Beyond}

Let us proceed to the subject of amplitudes which {\it vanish} in the soft limit.
A familiar example of this is the NGB, described by the general Lagrangian,
\eq{
{\cal L}_{{\rm NGB}} %& =  \sum_{n}\lambda_{2n}(\partial\phi)^{2n}\\
 & =  \frac{1}{2}(\partial\phi)^{2}+\frac{\lambda_{4}}{4!}(\partial\phi)^{4}+\frac{\lambda_{6}}{6!}(\partial\phi)^{6}+\ldots
}
Here the derivatively coupled structure of interactions is enforced by the nonlinearly realized shift symmetry of the NGB.
%$\phi\rightarrow\phi+{\rm constant}$.  
Since the field appears everywhere  with a derivative, all off-shell Feynman vertices are multilinear in each of the incoming momenta.  Consequently, the associated soft degree, $\sigma =1$, is trivially manifest at the level of the action.  Conversely, by {\it assuming} a soft degree of $\sigma =1$, we learn very little.  This constraint still permits an infinite space of theories parameterized by the coupling constants $\lambda_n$.

We can, however, up the ante to $\sigma=2$.  This implies that the amplitude vanishes at ${\cal O}(p)$ in the soft limit, which is only possible if there are nontrivial {\it cancellations} among contributions.  Concretely, for each $n$-particle amplitude, vanishing ${\cal O}(p)$ behavior requires that the coupling $\lambda_n$ of the $n$-particle vertex is specially related to the couplings at lower order.  Iterating this procedure \cite{Cheung:2014dqa}, one finds that all the $\lambda_n$ are determined in terms of the coupling constant of the leading interaction, $\lambda_4$.   The resulting coefficients telescope into the Taylor expansion of an {\it even simpler} Lagrangian,
\eq{
{\cal L}_{\textrm{DBI}} & =  \frac{1}{2}(\partial\phi)^{2}+ \frac{\lambda_{4}}{24}(\partial\phi)^{4}+\frac{\lambda_{4}^{2}}{144}(\partial\phi)^{6}+\frac{5\lambda_{4}^{3}}{3456}(\partial\phi)^{8}+\ldots\nonumber \\
 & =  -\frac{3}{\lambda_{4}}\sqrt{1-\frac{\lambda_{4}}{3}(\partial\phi)^{2}}+\textrm{constant},
}
which describes the scalar mode of Dirac-Born-Infeld (DBI) theory. In this theory the scalar labels  the coordinate
of a brane residing in an extra dimension. The scalar shift symmetry encodes translation symmetry in this direction.  Higher-dimensional rotations and boosts are manifested by an additional nonlinearly realized symmetry of the scalar which fixes the peculiar square root form of its interactions.  For these reasons we say that DBI theory has an ``enhanced soft limit": a soft degree which is higher than naively expected given the number of derivatives per field.  Achieving this special property requires {\it destructive interference} among contributions to the amplitude, which in turn dictates a rigid structure for interactions.

Why stop here?  It seems natural to further impose $\sigma=3$.  However, DBI is already defined by a single coupling constant, so setting ${\cal O}(p^2)$ soft contributions to zero will simply fix this parameter to zero, yielding a free theory.  So there is no interacting $\sigma=3$ theory with the power counting of one derivative per field.  DBI is as simple as it gets.

The concept of an enhanced soft limit is a generalization of the Adler zero \cite{Adler:1964um}, which is the observation that soft pions in the NLSM exhibit $\sigma=1$.  At face value this is surprising because the NLSM only has two derivatives per vertex\footnote{Earlier we proved that a single scalar with two-derivative interactions has a trivial S-matrix.  The NLSM evades this argument by introducing multiple flavors.}.  However, the NLSM encodes nonlinearly realized chiral symmetries which enforce these amplitude zeros in the soft limit.  Conversely, Susskind and Frye showed \cite{Susskind:1970gf} explicitly how the Adler zero can be reverse engineered in order to {\it derive} the structure of the NLSM.

Through a systematic generalization of the above procedure, one can carve out a theory space of possible Lorentz
invariant scalar EFTs \cite{Cheung:2016drk}.  The degree of symmetry in a theory correlates strongly with its soft behavior.  For the $n$-particle amplitude, the soft limit is enhanced when
\eq{
\textrm{derivatives per particle} = \frac{m}{n}  <  \sigma  = \textrm{soft degree}, \label{eq:enhanced_soft}
}
where $m$ is the number of derivatives in the $n$-particle vertex.  Imposing this criterion, we derive a special class of ``exceptional'' theories which exhibit the maximal enhanced soft limit permitted for a self-consistent and nontrivial S-matrix.  These exceptional theories include the NLSM ($\sigma=1$), DBI scalar theory ($\sigma=2$), and the special Galileon ($\sigma=3$).  The latter was first conjectured \cite{Cheung:2014dqa} and then later realized as an enhanced symmetry point \cite{Hinterbichler:2015pqa} of the original Galileon \cite{Nicolis:2008in} and as well as an output of the scattering equations \cite{Cachazo:2014xea}.   Perhaps unsurprisingly, these scalar EFTs are defined by a {\it single} coupling constant, consistent with the fact that they cannot be constrained further without becoming trivial.  In this sense, these theories are the scalar EFT analogs of YM theory and gravity!

While the soft limit fixes the {\it form} of interactions in a scalar EFT, what controls the {\it precise identity} of the nonlinearly realized symmetry?  For example, the pions of the NLSM are valued in $G/H$  corresponding to the breaking of a group $G$ down to a subgroup $H$.  Remarkably, this information is still accessible from the S-matrix through the commutator of soft limits \cite{Adler:1964um,Weinberg:1966kf,ArkaniHamed:2008gz}.

\subsection{Soft Recursion}

We can capitalize on the soft theorems to derive on-shell recursion relations for scalar EFTs.  Here the key insight is to relate the bothersome boundary term of the recursion to the infrared limit of the amplitude.

First, we need a momentum shift which can probe the soft limit of one or more external legs.  Here we will focus on the case of a $d$-dimensional shift \cite{Cheung:2015ota}, though there exist certain versions specialized to four dimensions \cite{Cheung:2015cba, Cachazo:2016njl}.  In particular, let us consider the soft momentum shift,
\eq{
p_{i} & \rightarrow p_{i}(1-za_{i}),
}
where $z$ is the usual deformation parameter and $a_{i}$
are constants. Since this shift is a rescaling of the momenta, it trivially preserves the on-shell
conditions.  However, total
momentum conservation is only guaranteed if
\eq{
\sum_{i}^{n}a_{i}p_{i} & =  0. \label{eq:soft_restriction}
}
  Unfortunately, this condition cannot always be satisfied in general.  If, for example, the total number of external particles is $n=d$, then the momentum vectors in \Eq{eq:soft_restriction} are linearly
independent and cannot sum to zero.  For $n=d+1$ legs the only solution to \Eq{eq:soft_restriction} is $a_{i}=1$, corresponding to a trivial rescaling of all the momentum.  Because the amplitude is homogenous in the momentum it also rescales, so this shift cannot probe any interesting kinematic regime of the amplitude. 
Thus, we conclude that a useful momentum shift requires that the number of particles exceed
\eq{
n & >  d+1.
}
  When this condition is satisfied, the $a_{i}$ can be made distinct so the shift probes the soft limit of leg $i$ as $z$ approaches $1/a_i$.  If the soft degree of the theory is $\sigma$, then
\eq{
\underset{z\rightarrow1/a_{i}}{\textrm{lim}}A(z) & \propto  (z-1/a_{i})^{\sigma}.
}
In other words, the amplitude has multiple zeroes of degree $\sigma$.

In order to construct an on-shell recursion relation it will be convenient to define a function $F(z)$ which vanishes on soft kinematics,
\eq{
F(z) & =  \prod_{i}^{n}(1-a_{i}z)^{\sigma},
}
where $F(0)=1$.  Repeating our procedure from before, we simply apply Cauchy's
theorem to the deformed amplitude with the slight twist of first dividing by $F(z)$, so
\eq{
A(z=0) & =  \frac{1}{2\pi i}\underset{z=0}{\oint}dz\;\frac{A(z)}{zF(z)}=-\sum_{I}\underset{z=z_{I}}{\textrm{Res}}\left[\frac{A(z)}{zF(z)}\right]+B_{\infty}.
}
The entire purpose of $F(z)$ is to improve the large $z$ behavior of the amplitude.  Naively, this comes at a cost---{\it new} denominator poles which enter at kinematic configurations which are not factorization channels and whose corresponding residues are completely unrelated to lower-point on-shell amplitudes.  However, all of these poles are {\it by construction} cancelled by the soft zeroes in $A(z)$, so no additional residues need be summed.  The only residues which contribute to the right-hand side of the recursion relation are the usual ones related to lower-point objects.

Last but not least is the question of the boundary term, $B_{\infty}$. For a general nonrenormalizable theory, the high energy behavior is poor and $A(z)$ will blow up at large $z$. The boundary term vanishes provided $F(z)$ grows even faster, so
\eq{
\underset{z\rightarrow\infty}{\textrm{lim}}\, \frac{A(z)}{F(z)} = 0\qquad\Rightarrow\qquad  B_{\infty}=0.
}
Since $F(z)$ is polynomial of degree $n\sigma$, it scales as $z^{n\sigma}$ at large $z$.  On the other hand, $A(z)$ scales at most as $z^m$ where $m$ is the total number of derivatives in the $n$-particle vertex.   So the boundary term vanishes when $\sigma > m/n$, coinciding precisely with the definition of the enhanced soft limit in \Eq{eq:enhanced_soft}.  In conclusion, we discover that all of the exceptional EFTs presented earlier---the NLSM, DBI theory, and the special Galileon---are constructible via on-shell recursion.

\section{Color-Kinematics Duality}

Up to now our efforts have been centered on reproducing results from QFT purely from on-shell methods.  While this is a noble aim in and of itself, the fact remains that nothing we have computed is formally beyond the capabilities of Feynman diagrams plus infinite time or RAM.  This is, however, besides the point---the true prize of the modern amplitudes program is the litany of {\it newly discovered structures}, long hidden in plain sight in theories more than a half a century old.

For the special case of ${\cal N}=4$ SYM, the fruits of this research program has been extraordinary.  On-shell tools have excavated a landscape of structures: hidden Yangian and dual conformal symmetries \cite{Drummond:2009fd,Drummond:2008vq,Berkovits:2008ic,Beisert:2008iq,Brandhuber:2008pf}, reformulations of the S-matrix in twistor space \cite{Penrose:1972ia,Hodges:1980hn,Witten:2003nn,Mason:2005zm,Boels:2006ir,AbouZeid:2006wu,Boels:2007qn,ArkaniHamed:2009si} and Grassmannian space \cite{ArkaniHamed:2009dn,Mason:2009qx,ArkaniHamed:2012nw}, as well as connections to abstract volumes \cite{Hodges:2009hk,ArkaniHamed:2010gg} like the amplituhedron \cite{Arkani-Hamed:2013jha}.

Perhaps less appreciated  outside the amplitudes community is that much of this progress also extends far beyond the case of ${\cal N}=4$ SYM.  In particular, color-kinematics duality \cite{Bern:2008qj} and the scattering equations \cite{Cachazo:2013hca,Cachazo:2013iea,Cachazo:2014xea} apply to an enormous class of theories, {\it e.g.}~including gravity and YM theory, but also EFTs like the NLSM!  While these structures are simple, provably true facts about the S-matrix, they are no less than sorcery from the perspective of the action.

A comprehensive survey of these new results would fill an entire course.   However, given the limitations of time, we focus here on the specific subject of color-kinematics duality, or Bern-Carrasco-Johansson (BCJ) duality \cite{Bern:2008qj}.  Epitomized by the word-algebraic equation, ``graviton = gluon${}^2$'', BCJ is a miraculous duality that interchanges the color and kinematic structures of the S-matrix.
Fundamentally, BCJ duality has two basic elements:
\begin{itemize}[leftmargin=0.5cm,rightmargin=0.5cm]

\item {\it Kinematic Algebra.}   Amplitudes can be rearranged so that their kinematic structures satisfy a kinematic analog of the Jacobi relations.  In this basis, color and kinematic structures are dual in the sense that they satisfy the exact same algebraic relations.  

\medskip

\item {\it Double Copy.}  Amplitudes in the dual form can be systematically ``squared'' to generate new amplitudes in {\it other} theories.  This construction generates a web of relations of the form ``graviton = gluon${}^2$'' and ``Galileon = pion${}^2$''.

\end{itemize}
\noindent BCJ duality is proven at tree-level \cite{Bern:2010yg} and vigorously checked
at higher loop \cite{Carrasco:2011mn,Carrasco:2012ca,Bjerrum-Bohr:2013iza,Boels:2013bi,Chiodaroli:2013upa,Bern:2013yya,Nohle:2013bfa}.  Remarkably, it is also a property of numerous QFTs.

\subsection{Color Structure}

Before moving forward it will be necessary to review the fundamentals of color in YM theory\footnote{Note that our discussion will also apply to {\it flavor} structures in the NLSM.}.   For additional details, we refer the reader to several excellent 
reviews \cite{Dixon:1996wi,Elvang:2015rqa,Carrasco:2015iwa} where this topic is covered in more depth.

For concreteness, let us consider YM theory with an $SU(N)$ gauge group.  The generators of the Lie algebra satisfy
\eq{
\textrm{Tr}(T_{a}T_{b}) =  \delta_{ab} \qquad \textrm{and} \qquad
[T_{a},T_{b}] & =  if_{abc}T_{c},
}
where the trace runs over the fundamental representation.  Here we have defined the structure constant,
\eq{
if_{abc} & =  \textrm{Tr}([T_{a},T_{b}]T_{c}),
}
which is cyclically invariant and antisymmetric in all of its indices.
We will make frequent use of the completeness relation for $SU(N)$,
\eq{
(T_{a})_{ij}(T_{a})_{kl} & =  \delta_{ik}\delta_{jl}-\frac{1}{N}\delta_{ij}\delta_{kl}. \label{eq:completeness}
}
To mitigate the proliferation of $\sqrt 2$ factors in amplitudes, our normalization convention differs slightly from the one typically used in textbooks.  

A gluon scattering amplitude is composed of kinematic factors multiplying color structures built from products of the structure constant, $f_{abc}$.  At tree-level, it is straightforward to reduce all such color structures into {\it single trace} objects.  As a concrete example, consider the Feynman diagram for $s$-channel gluon exchange in four-particle scattering.  The associated color factor is
\eq{
f_{a b e}f_{cde} & =  -\textrm{Tr}([T_{a},T_{b}]T_{e})\textrm{Tr}([T_{c},T_{d}]T_{e})\\
 & =  -\textrm{Tr}([T_{a},T_{b}][T_{c},T_{d}])+ {\cal O}(1/N).
}
By applying the completeness relation in \Eq{eq:completeness}, we can trivially rewrite this color structure in terms of single trace terms like $\textrm{Tr}(T_{a}T_{b}T_{c}T_{d})$,
plus $1/N$ corrections. Crucially, these
$1/N$ corrections actually vanish for tree-level gluon amplitudes.  To see why, realize that the $1/N$ terms serve only subtract the trace components of the generators. Adding back in these traces yields the completeness relation for $U(N)$, 
\eq{
(T_{a})_{ij}(T_{a})_{kl} & \overset{U(1)}{=}  \delta_{ik}\delta_{jl},
}
where all $1/N$ contributions disappear.  Physically, the extension from $SU(N)$ to $U(N)$ merely introduces a multiplet of $U(1)$ photons to the theory.  However, at the renormalizable level, photons decouple from all gluon interactions due to gauge invariance.  So all tree-level amplitudes involving
the photon will vanish and we can consistently drop the $1/N$ factors.

Every tree-level amplitude in YM theory has a ``color decomposition'' into a sum over single trace color structures times kinematic functions,
\eq{
A(1_{a_1}^{h_{1}}\cdots n_{a_{n}}^{h_{n}}) =  \sum_{\sigma\in S_{n}/\mathbb{Z}_{n}} \textrm{Tr}(T_{\sigma(a_{1})}\cdots T_{\sigma(a_{n})}) \, A(\sigma(1^{h_{1}})\cdots\sigma(n^{h_{n}})). \label{eq:color_decomposition}
}
The kinematic functions multiplying each color trace are gauge invariant, physical objects known as the ``color-ordered'' or ``partial''
amplitudes.  In a criminal abuse of notation, we will use the same symbol for both the full and color-ordered amplitude, distinguished only by the presence of color subscripts in the particle labels of the former.  

Color-ordered amplitudes are simpler than the full amplitudes.  The reason is that these objects have by definition an {\it intrinsic ordering} of the external legs, so their singularity structure is restricted.  In particular, factorization channels can only arise from {\it adjacent} sets of momenta, so for instance $1/s_{123}$ can appear but not $1/s_{135}$.

As an example, take the color decomposition of the four-particle gluon scattering amplitude,
 \eq{
A(1_{a}^{-}2_{b}^{-}3_{c}^{+}4_{d}^{+}) =& \phantom{{}+{}}  \textrm{Tr}(T_{a}T_{b}T_{c}T_{d}) \, A(1^{-}2^{-}3^{+}4^{+}) \\
 &   + \textrm{Tr}(T_{b}T_{c}T_{a}T_{d}) \, A(2^{-}3^{+}1^{-}4^{+})\\
 &   + \textrm{Tr}(T_{c}T_{a}T_{b}T_{d}) \, A(3^{+}1^{-}2^{-}4^{+})+\ldots \label{eq:colordecomp4}
 }
 Here the color-ordered amplitude is
 \eq{
A(1^{-}2^{-}3^{+}4^{+}) & =  \frac{\langle12\rangle^{4}}{\langle12\rangle\langle23\rangle\langle34\rangle\langle41\rangle}.
}
Since the color-ordered amplitude presumes a fixed ordering of the external legs, all of its singularities are in adjacent momenta.   The same is true of the Park-Taylor formula in  \Eq{eq:PT}.

 The color decomposition in \Eq{eq:color_decomposition} is a function of $(n-1)!$ trace structures.  However, there are only $(n-2)!$ {\it independent} color structures spanned by a basis of the form 
\eq{
[\tau_{\sigma(a_2)} \cdot \tau_{\sigma(a_3)} \cdots \tau_{\sigma(a_{n-2})} \cdot \tau_{\sigma(a_{n-1})}]_{a_1 a_n}
}
where legs 1 and $n$ have been fixed and $\sigma \in S_{n-2}$ is any permutation of the inner $n-2$ legs. Here the products represent matrix multiplication in the adjoint representation, so $if_{abc} = [\tau_b]_{ac}$.  Any possible string of structure constants can be expressed as a linear combination of the above objects via the Jacobi identities.   This basis is sometimes called the ``half-ladder'' basis, in reference to the structure of color index contractions when expressed in terms of $f_{abc}$ factors.

Using the half-ladder basis one can construct an {\it even more minimal} color decomposition of the YM amplitude,
\eq{
&A(1_{a_1}^{h_{1}}\cdots n_{a_{n}}^{h_{n}}) =  \sum_{\sigma\in S_{n-2}} [\tau_{\sigma(a_2)} \cdots  \tau_{\sigma(a_{n-1})}]_{a_1 a_n} \, A(1^{h_{1}}\sigma(2^{h_{2}})\cdots\sigma(n^{h_{n-1}}) n^{h_{n}}). \label{eq:DDM} 
}
known as the DDM basis \cite{DelDuca:1999rs}.  Note that the number of independent color-ordered amplitudes has decreased from $(n-1)!$ to $(n-2)!$, consistent with the Kleiss-Kuijf relations \cite{Kleiss:1988ne}.  As an example, the DDM decomposition of the four-particle YM amplitude is
 \eq{
A(1_{a}^{-}2_{b}^{-}3_{c}^{+}4_{d}^{+}) &=[\tau_{b} \cdot  \tau_{c}]_{ad} \, A(1^{-}2^{-}3^{+}4^{+}) + [\tau_{c} \cdot  \tau_{b}]_{ad} \, A(1^{-}3^{+} 2^{-}4^{+}),
 }
 which is even more compact than \Eq{eq:colordecomp4}.

\subsection{A Simple Example}

Armed with a basic understanding of color management, we are finally equipped to discuss color-kinematics duality.  To be concrete, consider the simplest example of the four-particle amplitude in YM theory.  As we saw earlier, this object appears with three possible color structures,
\eq{
c_{s} & =  f_{abe}f_{cde}\\
c_{t} & =  f_{bce}f_{ade}\\
c_{u} & =  f_{cae}f_{bde},
}
corresponding to the $s$-, $t$-, and $u$-channel exchange diagrams.  Previously, we showed how factorization implies the Jacobi identity,
$c_{s}+c_{t}+c_{u} =  0$.  Note the similarity of this equation to the {\it kinematic} relation, $s+t+u=0$. As we will soon see, this is not mere coincidence!

Starting from {\it any} representation of the four-particle scattering amplitude, one can always rearrange terms into to the form
\eq{
A(1_{a}2_{b}3_{c}4_{d}) & =  \frac{c_{s}n_{s}}{s}+\frac{c_{t}n_{t}}{t}+\frac{c_{u}n_{u}}{u}. \label{eq:BCJform}
}
While the Feynman diagram representation of the amplitude obviously has cubic exchange diagrams in this form, the same is not true for the quartic vertex contributions.  However, these contact terms can be massaged into a cubic form simply by canceling denominator poles with additional numerator factors.  That is, any quartic vertex can be recast as a cubic exchange diagram by inserting the equation,
\eq{
1 & =  \frac{s}{s}=\frac{t}{t}=\frac{u}{u},
}
whose proof we leave as an exercise for the reader.  

Of course, any such rearrangement of contact terms will not be unique, but this is exactly the point.  By exploiting this freedom, BCJ conjectured that there {\it always} exists a form of the amplitude in which the numerators satisfy the Jacobi identity,
\eq{
n_{s}+n_{t}+n_{u} & =  0.
}
Since the BCJ numerators satisfy the same identities as the color factors, it is said that this representation manifests color-kinematics duality.
That a BCJ form should exist may seem like a miracle, but we have at our disposal an immense redundancy.  That is, for various choices of gauge and field basis, the distribution of terms between the $s$-, $t$-, and $u$-channel topologies will change.   Under this broad set of ``generalized gauge transformations'', the numerators of the amplitude can {\it only} shift by
\eq{
n_{s} & \rightarrow  n_{s}+\alpha s\\
n_{t} & \rightarrow  n_{t}+\alpha t\\
n_{u} & \rightarrow  n_{u}+\alpha u,
}
so that the physical amplitude remains invariant,
\eq{
A_4 & \rightarrow A_4+\alpha \left(c_{s}+c_{t}+c_{u}\right)=A_4,
}
by the color Jacobi identities.  Here $\alpha$ is an arbitrary, possibly non-local function of the external kinematics parameterizing the space of generalized gauge transformations.

Because the four-particle amplitude is so simple it is easy to derive explicit expressions for the BCJ numerators.  From our earlier discussion we argued that the four-particle amplitudes takes the form
\eq{
A(1_{a}^-2_{b}^-3_{c}^+4_{d}^+)  & =  \langle12\rangle^{2}[34]^{2}\left(\frac{c_{st}}{st}+\frac{c_{tu}}{tu}+\frac{c_{us}}{us}\right), \label{eq:YM4compare}
}
where $c_{st}$, $c_{tu}$, and $c_{us}$ are dimensionless constants. Proper factorization implies that these constants are related to the color structures by
\eq{
c_{st}-c_{us} & =  c_{s}\\
c_{tu}-c_{st} & =  c_{t}\\
c_{us}-c_{tu} & =  c_{u}.
}
Comparing \Eq{eq:BCJform} and \Eq{eq:YM4compare}, we obtain
\eq{
A(1_{a}^-2_{b}^-3_{c}^+4_{d}^+)  & =  \frac{(c_{st}-c_{us})n_{s}}{s}+\frac{(c_{tu}-c_{st})n_{t}}{t}+\frac{(c_{us}-c_{tu})n_{u}}{u} \nonumber\\
 & =  \frac{c_{st}(tn_{s}-sn_{t})}{st}+\frac{c_{tu}(un_{t}-tn_{u})}{tu}+\frac{c_{us}(sn_{u}-un_{s})}{us}.
}
Matching term by term, we obtain  three equations for the BCJ  numerators,
\eq{
tn_{s}-sn_{t} & =  \langle12\rangle^{2}[34]^{2}\\
un_{t}-tn_{u} & =  \langle12\rangle^{2}[34]^{2}\\
sn_{u}-un_{s} & =  \langle12\rangle^{2}[34]^{2},
}
which as a matrix equation becomes
\eq{
\left(\begin{array}{ccc}
t & -s & 0\\
0 & u & -t\\
-u & 0 & s
\end{array}\right)\left(\begin{array}{c}
n_{s}\\
n_{t}\\
n_{u}
\end{array}\right) & =  \langle12\rangle^{2}[34]^{2}\left(\begin{array}{c}
1\\
1\\
1
\end{array}\right).
}
The above matrix is not invertible because it has vanishing determinant.  This is not surprising because $(n_{s},n_{t},n_{u})=(s,t,u)$ is a zero eigenvector
of the matrix, corresponding to generalized gauge invariance. No matter
though\textemdash the fact that the matrix is not invertible simply implies that there is a family of solutions.  Solving for $n_s$ and $n_t$ in terms of $n_u$, we find
\eq{
n_{s} & =  -\frac{\langle12\rangle^{2}[34]^{2}}{u}+\frac{n_{u}s}{u}\\
n_{t} & =  \frac{\langle12\rangle^{2}[34]^{2}}{u}+\frac{n_{u}t}{u},
%\\
%n_{u} & =  \textrm{anything},
}
for arbitrary $n_{u}$.  Remarkably, the kinematic Jacobi identity, $n_{s} + n_{t}+n_{u}=0$, is satisfied for any $n_u$ as a direct consequence of $s+t+u=0$.   Thus, {\it any} representation of the four-particle amplitude provides a viable set of BCJ numerators.  This is a very special property of four-particle kinematics.

As a double check, one can compute the color-ordered amplitude, 
\eq{
A(1^{-}2^{-}3^{+}4^{+}) & = -\frac{n_{s}}{s}+\frac{n_{t}}{t}=\frac{\langle12\rangle^{2}[34]^{2}}{u}\left(\frac{1}{s}+\frac{1}{t}\right)\\
 & =  -\frac{\langle12\rangle^{2}[34]^{2}}{st} =  \frac{\langle12\rangle^{4}}{\langle12\rangle\langle23\rangle\langle34\rangle\langle41\rangle},
}
which correctly matches the Park-Taylor formula.  Note the various signs entering through the color factors.

Because the BCJ numerators satisfy the Jacobi identities, we can think of the latter as {\it defining} relations for the former.  In this sense there is little difference, at least algebraically, between the color and kinematic factors in the amplitude.  Thus, it is natural to simply {\it interchange} the color factors with kinematic factors in various ways.  
For instance, by substituting in color for kinematics, we obtain
\eq{
A(1234) & =  \frac{c_{s}^{2}}{s}+\frac{c_{t}^{2}}{t}+\frac{c_{u}^{2}}{u},
}
which is the amplitude for a theory of scalars coupled via trilinear interactions.  The form of the color structures indicates that each scalar has two independent adjoint color indices.  For obvious reasons this theory is known as the bi-adjoint scalar (BS) theory.
Alternatively, we can substitute kinematics for color, yielding
\eq{
A(1^{--}2^{--}3^{++}4^{++}) & =  \frac{n_{s}^{2}}{s}+\frac{n_{t}^{2}}{t}+\frac{n_{u}^{2}}{u}\\
 & =  \frac{\langle12\rangle^{4}[34]^{4}(s+t)+n_{u}^{2}st(s+t+u)}{stu^{2}}\\
 & =  -\frac{\langle12\rangle^{4}[34]^{4}}{stu},
}
which is precisely the four-particle graviton amplitude!    This construction is referred to as the ``double copy'' for obvious reasons.  However, it is a bit of a misnomer because one need not replace color factors with the kinematic factors appearing in the same amplitude.  One can instead mix and match color and kinematic factors from {\it different} amplitudes, yielding asymmetric copies which turn out to be equally sensible scattering amplitudes.

\subsection{Universality of the Double Copy}

The extraordinary structures introduced in the previous sections are apparently ubiquitous.  Not only do these concepts extend to higher-point tree- and loop-level amplitudes, they also generalize across a huge range of theories.  That is, if you can think of a Lorentz invariant
theory with a single coupling constant and a name, chances are it lies on one or another side of a double copy relation. 

BCJ duality holds for the $n$-particle tree-level YM amplitude, which can always be recast into the form
\eq{
A_{{\rm YM}} & =  \sum_{i}\frac{c_{i}n_{i}}{d_{i}},
}
where the sum runs over all cubic topologies and the $d_i$ are the associated products of propagator denominators.  As before, all quartic and higher vertices are trivially recast as cubic diagrams by multiplying and then dividing by the appropriate propagator factors.    

Within the full set of cubic topologies, one can always pick out three which differ only by a single propagator factor.  This is a higher-point generalization of the classification of $s$-, $t$-, and $u$-channel diagrams.  BCJ duality implies that the numerators can always be chosen so that for {\it every} triplet of topologies $i,j,k$ the associated color and kinematic factors satisfy
\eq{
c_{i}+c_{j}+c_{k}  =  0 \qquad \textrm{and} \qquad
n_{i}+n_{j}+n_{k}  =  0.
}
In this BCJ form we can substitute kinematics for color to obtain the amplitude of BS theory,
\eq{
A_{{\rm BS}} & =  \sum_{i}\frac{c_{i}^{2}}{d_{i}},
}
or substitute color for kinematics to obtain the graviton amplitude,
\eq{
A_{{\rm G}} & =  \sum_{i}\frac{n_{i}^{2}}{d_{i}}.
}
Hence, the double copy construction is a rather elegant method for leapfrogging from gauge theory directly to gravity.  Moreover, the BCJ construction applies in {\it general} dimensions, so it does not hinge on any special properties of four-dimensional kinematics.

As noted, BCJ duality extends beyond the case of gauge theory and gravity.  For instance, one can express the amplitudes of the NLSM as
\eq{
A_{{\rm NLSM}} & =  \sum_{i}\frac{c_{i}r_{i}}{d_{i}},
}
where $r_i$ are kinematic numerators satisfying $r_i + r_j+ r_k=0$, and the $c_i$ are the same objects defined earlier except interpreted physically as {\it flavor} structures.  Again substituting color for kinematics, we obtain
\eq{
A_{{\rm SG}} & =  \sum_{i}\frac{r_{i}^2}{d_{i}},
}
which is the amplitude for the special Galileon.  Taking an asymmetric product of the YM and NLSM numerators, we obtain
\eq{
A_{{\rm BI}} & =  \sum_{i}\frac{n_i r_{i}}{d_{i}},
}
which is the amplitude of BI theory.  These relations are concisely summarized in the ``multiplication table'' of BCJ double copies depicted in Fig.~\ref{fig:table}.   

The universality of the double copy is not yet fully understood.  A partial explanation can be found in the scattering equations of Cachazo, He, and Yuan \cite{Cachazo:2013hca,Cachazo:2013iea,Cachazo:2014xea}, who proposed an extraordinarily compact construction for the S-matrices of {\it all} the theories in Fig.~\ref{fig:table}.  That this is even possible is a manifestation of a hidden unity among these theories, instantiated by a set of simple relations \cite{Cheung:2017ems} which extract all their S-matrices from that of gravity.

\begin{figure}[t]
\begin{center}
\begin{tabular}{c||c|c|c|}
 		& BS & NLSM & YM  \\ \hline\hline
BS  		& BS & NLSM & YM \\ \hline
NLSM  	& NLSM & SG & BI \\ \hline
YM    	& YM & BI & G \\ \hline
\end{tabular}
\end{center}
\caption{Multiplication table of QFTs, including bi-adjoint scalar (BS) theory, the nonlinear sigma model (NLSM), Yang-Mills (YM) theory, the special Galileon (SG), Born-Infeld (BI) theory, and gravity (G).} \label{fig:table}
\end{figure}

Last but not least, the double copy also generalizes straightforwardly to the case of supersymmetric theories.  For instance, the product of ${\cal N}=4$ SYM with a second copy of ${\cal N}=0,1,2$, or 4 SYM is simply ${\cal N}=4,5,6$, or 8 supergravity!  Other combinations of gauge theories produce non-pure supergravities with additional matter.  These relations have been critical for alleviating some of the immense technical challenges in studying the ultraviolet properties of ${\cal N}=8$ supergravity \cite{Bern:1998ug,Bern:2012gh}. 

\subsection{Beyond Scattering Amplitudes}

Color-kinematics duality is especially noteworthy because it quite possibly extends {\it beyond} the context of the perturbative S-matrix.   In particular, an apparent vestige of the duality has been discovered in {\it classical} solutions in gauge theory and gravity.  In its simplest incarnation, ``black hole = charge${}^2$'' \cite{Monteiro:2014cda,Monteiro:2015bna,Ridgway:2015fdl,Luna:2016due}, the Schwarzschild solution of general relativity is shown to be dual to the Coulomb potential of electrodynamics.  Since then, similar double copy  structures have also been found in classical radiation  \cite{Goldberger:2016iau,Goldberger:2017frp} and other curved backgrounds \cite{Luna:2015paa,Adamo:2017nia}.   

In hindsight, a classical manifestation of the double copy is not unexpected---after all, tree-level amplitudes are {\it literally} the solutions to classical equations of motion initialized by plane wave perturbations at infinity \cite{Boulware:1968zz}.  Indeed, Duff \cite{Duff:1973zz} showed long ago how to construct the Schwarzschild solution by an appropriate convolution of perturbative scattering amplitudes.   Nevertheless, the above examples elevate the notion of a nonperturbative double copy from hallucinatory to tantalizingly possible.

\subsection{Physical Origins}

While no one disputes that color-kinematics duality {\it is} true, the question remains, {\it why} it is true?  That is, despite myriad mathematical proofs of its veracity, the {\it physical} principle behind the duality remains elusive.   Where in the textbook derivation of QFT are the secret instructions to rearrange diagrams into cubic structures satisfying kinematic Jacobi identities and so on and so forth?

It is tempting to contemplate an explanation for color-kinematics duality purely within the confines of standard QFT---ideally, from an explicit action.  In the best of all possible worlds, such an action would manifest  {\it all} the symmetries of the double-copy construction: {\it i}) a parity exchanging color and kinematic algebras, and {\it ii}) a pair of independent Lorentz symmetries.  The latter is a trivial consequence of the fact that the double copy amplitudes are built from {\it separately} Lorentz invariant numerators.  

Recent years have seen incremental progress towards an action-level understanding of these structures.  For instance, Monteiro and O'Connell  \cite{Monteiro:2011pc} beautifully demonstrated how the self-dual sector of YM theory squares into self-dual gravity.  In this formulation the cubic Feynman vertices are the structure constants of the kinematic algebra.  Unfortunately, this has yet to generalized to YM beyond the self-dual limit or outside of four dimensions. 

 More recently, color-kinematics duality has been manifested explicitly in a cubic reformulation of the NLSM action \cite{Cheung:2016prv}.  The associated Feynman diagrams automatically produce expressions compliant with the BCJ form, while the Jacobi identities arise from an off-shell symmetry closely connected to Lorentz invariance.  Squaring the {\it action} then yields a new action for the special Galileon which manifests a twofold Lorentz invariance.  While it is also possible to rewrite the Einstein-Hilbert action in a way that manifests a twofold Lorentz invariance \cite{Cheung:2016say,Cheung:2017kzx}, the connection to color-kinematics duality remains murky.

\section{Outlook}

Upon the discovery of the muon, Isidor Rabi famously quipped ``who ordered that?"  Appropriate to the times, those words affirm an implicit expectation that there are many ways that the world {\it could} have been, but only one way that it {\it is}.  In one sense this is true.  After all, the standard model exhibits a baroque pattern of masses and couplings.  

 On the other hand, the on-shell formulation of QFT demonstrates how dynamical structures like the {\it form} of interactions in gravity, gauge theory, and effective field theories are fixed {\it uniquely and directly} by physical principles.  
In this respect, nature's menu is prix fixe.   By relinquishing the action principle---together with its perks and pitfalls---we are finally able to see some of the underlying structures which have lain dormant despite decades of study.   There is still much to understand!

\section*{Acknowledgements}

We are grateful to Rouven Essig and Ian Low for the invitation to speak at TASI 2016.  Also, we would like to thank to Zvi Bern, Grant Remmen, Chia-Hsien Shen, Congkao Wen, Mark Wise, and especially Jake Bourjaily for comments on the manuscript.
CC is supported by a Sloan Research Fellowship and a DOE Early Career Award under Grant No. DE-SC0010255.

\section*{Further Reading} There are a many of excellent introductory treatments of scattering amplitudes in review articles and textbooks.  A very partial list includes:

\begin{itemize}[leftmargin=0.5cm,rightmargin=0.5cm]

 \item  {\it ``Gauge and Gravity Amplitudes Relations''} \cite{Carrasco:2015iwa}, by Carrasco
\medskip

\item {\it ``Calculating Amplitudes Efficiently''} \cite{Dixon:1996wi}, {\it ``Scattering Amplitudes: the Most Perfect Microscopic Structures in the Universe''} \cite{Dixon:2011xs},
 and {\it ``A Brief Introduction to Modern Amplitudes Methods} \cite{Dixon:2013uaa},  by  Dixon
\medskip

 \item  {\it ``Scattering Amplitudes''} \cite{Elvang:2013cua} and {\it ``Scattering Amplitudes in Gauge Theory and Gravity''} \cite{Elvang:2015rqa},  by Elvang and Huang  
\medskip

 \item  {\it ``Scattering Amplitudes in Gauge Theories''} \cite{Henn:2014yza}, by Henn and Plefka
\medskip

\item {\it ``Quantum Field Theory and the Standard Model''} \cite{Schwartz:2013pla}, by Schwartz 
\medskip

\item {\it ``Quantum Field Theory in a Nutshell''} \cite{Zee:2003mt}, by Zee

\end{itemize}

\appendix

\begin{appendices}

\section{Counting Feynman Diagrams}

\label{app:count}

For an $n$-particle amplitude the corresponding number of Feynman diagrams $c_n$ typically grows quite rapidly with $n$.  In this appendix we derive explicit formulae for $c_n$ in a theory with arbitrary interactions.  

\medskip

\noindent {\it Cubic Scalar Theory.} To start, consider the case of $\phi^3$ theory.  
We introduce ``toy'' equations of motion in which the d'Alembertian and the cubic coupling have been set to one and $\phi$ is simply a function, so
\eq{
\phi = J+\frac{\phi^2}{2!} . \label{eq:phi3EOM}
}
The perturbative solution to the equations of motion in the presence of a source is in one-to-one correspondence with the generating functional of connected tree amplitudes \cite{Boulware:1968zz}.  Since the toy equations of motion effectively set all propagators and Feynman vertices to one, the corresponding solution simply counts the number of diagrams. So $c_n$ is extracted from the numerical coefficient of the $J^{n-1}$ term in the solution.  Solving \Eq{eq:phi3EOM}, we obtain 
\eq{
\phi(J) &= \phi(J) = 1 - \sqrt{1 - 2 J} \\
&= J + (1) \frac{J^2}{2!}  + (3) \frac{J^3}{3!}  + (15)\frac{J^4}{4!}  + (105)\frac{J^5}{5!}  + \ldots,
}
where we have chosen the root satisfying $\phi(0)=0$.  We thus correctly conclude that the number of cubic Feynman diagrams is $c_n = (2n-5)!!$.

\medskip

\noindent {\it Yang-Mills Theory.} The same logic applies to theories with higher-degree interaction vertices.  For example, YM theory has both cubic and quartic vertices, so the toy equations of motion are
\eq{
A = J+\frac{A^2}{2!} + \frac{A^3}{3!}  \label{eq:phi3EOM}.
}
Like before, we choose the root with $A(0)=0$, so
\eq{
A(J)= & -1+\frac{3
   \left(1-i \sqrt{3}\right)}{2
   \sqrt[3]{\sqrt{9 J^2+24 J-11}+3 J+4}} \\ 
& +\frac{1}{2} \left(1+i \sqrt{3}\right)
   \sqrt[3]{\sqrt{9 J^2+24 J-11}+3 J+4} \\
= & \; J + (1) \frac{J^2}{2!}  + (4) \frac{J^3}{3!}  + (25)\frac{J^4}{4!}  + (220)\frac{J^5}{5!}  + \ldots,
}
which is in agreement with the counting from known results.

\medskip

\noindent {\it Gravity.} Last but not least is the case of gravity, whose toy equations of motion extend to infinite order, 
\eq{
h = J+ \frac{h^2} {2!} + \frac{h^3} {3!} + \frac{h^4} {4!} +\ldots = J+ e^h - 1 - h ,
}
which has a semi-closed form solution,
\eq{
h(J) &= \frac{J-1}{2} -  W\left( -e^{\frac{J-1}{2}}/2 \right) \\
&= J + (1) \frac{J^2}{2!}  + (4) \frac{J^3}{3!}  + (26)\frac{J^4}{4!}  + (236)\frac{J^5}{5!}  + \ldots,
}
where $W$ is the product logarithm function.

%\end{appendix}

\bigskip

\section{Basis of Mandelstam Invariants}

\label{app:mandelstam}

In general dimensions, any $n$-particle amplitude can be expressed as a function of the Mandelstam invariants,
\eq{
\{ p_i p_j \} , \qquad  \{ p_i e_j \}, \qquad  \{ e_i e_j \},
}
for all $1\leq i,j \leq n$ where $i\neq j$.   While the $\{ e_i e_j \}$ are all independent variables, the same cannot be said for $\{ p_i p_j \}$ and $\{ p_i e_j \}$ because of momentum conservation. On the other hand, one can always prune down to a linearly independent basis.  For instance by eliminating $p_n$ via momentum conservation, we can effectively drop all Mandelstam invariants of the form $p_n p_i $ and $p_n e_i$.
%\eq{
%0=  \sum_{j=1}^n p_i p_j  \qquad \textrm{for} \qquad 1\leq i <n. \label{eq:momentum_eliminate}
%}
Furthermore, the on-shell condition for $p_n$ implies
\eq{
0=  \sum_{i}^{n-1}\sum_{j}^{n-1} p_i p_j  , \label{eq:onshell_eliminate}
}
while the transverse condition on $e_n$ implies
\eq{
0=  \sum_{i}^{n-1} p_i e_n \label{eq:transverse_eliminate}.
}
Altogether, these eliminate $n$ of the $n(n-1)/2$ variables in $\{ p_i p_j \}$ and $n$ of the $n(n-1)$ variables in $\{ p_i e_j \}$.  Obviously leaves an immense freedom in choosing which variables to eliminate.  However, any full set of linearly independent Mandelstam invariants takes the form
\eq{
%I^{(n)} = 
\{ p_i p_j \}_{\frac{n(n-3)}{2}}, \qquad  \{ p_i e_j \}_{\scriptscriptstyle n(n-2)}, \qquad \{ e_i e_j \}_{\frac{n(n-1)}{2}}  ,
}
where the subscripts denote the number of elements in each set.

Depending on the circumstance, different forms of this basis may be appropriate.  For example, if one is interested in an $n$-particle amplitude with adjacent factorization channels, then for odd $n$, a convenient basis is
\eq{
\{ p_i p_j \}_{\frac{n(n-3)}{2}}
= \left\{\right. 
& s_{12}, s_{123}, s_{1234},\ldots, s_{12\cdots \frac{n-1}{2}}
 \left.\right\}  + \textrm{cyclic perm},
}
where we have defined the generalized Mandelstam invariant,
\eq{
s_{i_1 i_2 \cdots i_ m} = (p_{i_1} + p_{i_2} + \ldots + p_{i_m})^2.
}
For the case of even $n$, a similar basis exists except that the set of $n$ variables of the form $s_{12\cdots \frac{n-1}{2}}$ is replaced with $n/2$ variables of the form $s_{12\cdots \frac{n}{2}}$.

\end{appendices}

%\end{appendix}

\clearpage
\pagestyle{empty}

%\newpage

\bibliographystyle{JHEP}
\bibliography{TASIbib}

\end{document}